\documentclass[varenna,nocopyright]{cimento}

\usepackage{graphicx}  
\usepackage{braket} 
\usepackage{cite}
\usepackage{bm}
\usepackage{amsmath}
\usepackage{amssymb}
\usepackage{amsfonts}
\usepackage{upgreek}
\usepackage{xspace}
\usepackage{siunitx}
\usepackage{cleveref}
\usepackage{dsfont}

\newcommand*\diff{\mathop{}\!\mathrm{d}}

\newcommand{\keff}{k_{\mathrm{eff}}}
\newcommand{\keffi}[1]{k_{\mathrm{eff},#1}}
\newcommand{\keffvc}{{\bf k}_{\mathrm{eff}}}

\newcommand{\tsep}{\tau_{\mathrm{sep}}}

\newcommand{\Isat}{I_{\mathrm{sat}}}

\newcommand{\quot}[1]{``#1''}

\newcommand{\Rbb}{\textsuperscript{85}Rb}
\newcommand{\Rb}{\textsuperscript{87}Rb}
\newcommand{\K}{\textsuperscript{39}K}

\newcommand{\Oeg}{\Omega_{\mathrm{eg}}}
\newcommand{\Oeff}{\Omega_{\mathrm{eff}}}
\newcommand{\orec}{\omega_{\mathrm{rec}}}
\newcommand{\oeg}{\omega_{\mathrm{eg}}}

\newcommand{\EotDelta}{$\Delta\eta$}
\newcommand{\Eotsigma}{$\delta\eta$}
\newcommand{\spaceminus}{$\thinspace -\thinspace$}
\newcommand{\Q}[1]{$Q^{'1}_{\text{#1}}$}
\newcommand{\QQ}[1]{$Q^{'2}_{\text{#1}}$}
\newcommand{\fbminus}[1]{\mbox{$f_{\beta^{e+p-n}_{\text{#1}}}$}}

\newcommand{\fbbarminus}[1]{\mbox{$f_{\beta^{\bar{e}+\bar{p}-\bar{n}}_{\text{#1}}}$}}
\newcommand{\fbplus}[1]{\mbox{$f_{\beta^{e+p+n}_{\text{#1}}}$}}

\newcommand{\fbbarplus}[1]{\mbox{$f_{\beta^{\bar{e}+\bar{p}+\bar{n}}_{\text{#1}}}$}}

\newcommand{\RN}[1]{\uppercase\expandafter{\romannumeral#1}}

\newcommand{\stat}{5.4\cdot 10^{-7}}
\newcommand{\etav}{(0.3\pm 5.4)\cdot 10^{-7}}

\usepackage{ifthen} 

\def\schl#1{\ifthenelse{\equal{#1}{i}}{\widetilde{\imath}}{
		\ifthenelse{\equal{#1}{j}}{\widetilde{\jmath}}{\widetilde{#1}} } } 

\def\be#1{\begin{equation} \label{#1}}  \def\ee{\end{equation}}
\def\bea#1{\begin{eqnarray} \label{#1}}  \def\eea{\end{eqnarray}} 

   \let\b=\beta     
  \let\o=\omega    
     
\let\vrho=\varrho

\let\vec\mathbf

\def\pueb#1#2{\hspace*{-#1mm}{\buildrel{\hspace*{#1mm}
			\hspace*{#1mm} _\rightharpoonup}\over{#2}}\hspace*{-#1mm}
	\hspace*{-#1mm}\hspace*{-.2mm}}
\def\gpueb#1#2{\hspace*{-#1mm}\hspace*{.4mm}{\buildrel{
			\hspace*{#1mm}\hspace*{#1mm}{\displaystyle _\rightharpoonup}}
		\over{#2}}\hspace*{-#1mm}\hspace*{-#1mm}\hspace*{.4mm}}
\def\vc#1{
	\ifthenelse{\equal{#1}{f} \or \equal{#1}{d}}{\pueb{.4}{#1}}{
		\ifthenelse{\equal{#1}{e} \or \equal{#1}{k} \or \equal{#1}{l} 
			\or \equal{#1}{r} \or \equal{#1}{s} \or \equal{#1}{t} 
			\or \equal{#1}{\b} \or \equal{#1}{\beta} \or \equal{#1}{\ell} 
			\or \equal{#1}{\phi} \or \equal{#1}{\vrho} 
			\or \equal{#1}{\varrho}}{\pueb{.2}{#1}}{   
			\ifthenelse{\equal{#1}{j}}{\pueb{.4}{\jmath}}{
				\ifthenelse{\equal{#1}{i}}{\pueb{.4}{\imath}}{
					\ifthenelse{\equal{#1}{m} \or \equal{#1}{w} \or \equal{#1}{\o}
						\or \equal{#1}{\omega}}{\gpueb{0}{#1}}{
						\ifthenelse{\equal{#1}{A} \or \equal{#1}{R}}{\gpueb{.2}{#1}}{
							\ifthenelse{\equal{#1}{B} \or \equal{#1}{C} \or \equal{#1}{D} 
								\or \equal{#1}{E} \or \equal{#1}{F} \or \equal{#1}{G} \or \equal{#1}{H} 
								\or \equal{#1}{I} \or \equal{#1}{J} \or \equal{#1}{K} \or \equal{#1}{L} 
								\or \equal{#1}{M} \or \equal{#1}{N} \or \equal{#1}{O} \or \equal{#1}{P} 
								\or \equal{#1}{Q} \or \equal{#1}{S} \or \equal{#1}{T} \or \equal{#1}{U} 
								\or \equal{#1}{V} \or \equal{#1}{W} \or \equal{#1}{X} \or \equal{#1}{Y} 
								\or \equal{#1}{Z}}{\gpueb{.4}{#1}}{\pueb{0}{#1}}} }}} }} }

\title{Atom interferometry and its applications}

\author{S. Abend, M. Gersemann, C. Schubert, D. Schlippert, E.~M. Rasel}
\institute{Institut f\"ur Quantenoptik, Leibniz Universit\"at Hannover - Welfengarten 1, D-30167 Hannover, Germany}
\author{M. Zimmermann, M. A. Efremov, A. Roura}
\institute{Institut f\"ur Quantenphysik, Universit\"at Ulm - Albert-Einstein-Allee 11, D-89081 Ulm, Germany}
\author{F.~A. Narducci}
\institute{Naval Postgraduate School, Department of Physics - Monterey, CA 93943, USA}
\author{W.~P. Schleich}
\institute{Institut f\"ur Quantenphysik and Center for Integrated Quantum Science and Technology (IQ$^{ST}$), Universit\"at Ulm - 
Albert-Einstein-Allee 11, D-89081 Ulm, Germany\\
Institute for Quantum Science and Engineering (IQSE), Texas A$\&$M AgriLife Research, Hagler Institute 
for Advanced Study at Texas A$\&$M University, Department of Physics and Astronomy, 
Texas A$\&$M University - College Station, TX 77843-4242, USA}

\PACSes{\PACSit{03.75.DG}{Atom and neutron interferometry}
\PACSit{37.25.+k}{Atom interferometry techniques}
\PACSit{42.50.-p}{Quantum optics}
\PACSit{04.80.Cc}{Experimental tests of gravitational theories}
}

\shortauthor{S. Abend et al.}

\begin{document}
\maketitle
\vspace*{-2cm}

\begin{abstract}
We provide an introduction into the field of atom optics and review our work on interferometry with cold atoms, and in particular with Bose-Einstein condensates. Here we emphasize applications of atom interferometry with sources of this kind. We discuss tests of the equivalence principle, a quantum tiltmeter, and a gravimeter.
\end{abstract}

\newpage


\section{Introduction}
Based on the pioneering work~\cite{Kasevich91PRL,Kasevich92APB} by Mark~Kasevich and Steve~Chu starting in 1991, light-pulse atom interferometry has grown into an extremely successful tool for precision measurements. Indeed, ground-breaking experiments have been performed in the fields of inertial sensing and tests of the foundations of physics. Inertial sensing covers measurements of the local gravitational acceleration~\cite{Peters99Nature,Peters01Metrologia,PhysRevA.86.043630,Malossi10PRA,0256-307X-28-1-013701,Schmidt2011,Zhou11GRG,McGuinness12APL,Debs11PRA,Charriere12PRA,Altin13NJP,
Bidel13APL,Hauth13APB,PhysRevA.88.023614,Andia13PRA,Bonnin13PRA,Biedermann15PRA,Baryshev2015,1402-4896-91-5-053006,LeGouet08APB,Hu13PRA}, or rotations, for example of the Earth~\cite{Gustavson97PRL,Gauguet09PRA,Stockton11PRL,Berg15PRL,Dutta16PRL,PhysRevLett.97.240801,PhysRevLett.99.173201}, as well as gravity gradiometry~\cite{McGuirk02PRA,Lamporesi08PRL}. The present lecture notes aim at providing an introduction into,  and an overview over this rapidly moving field. Moreover, our article complements the corresponding lectures by Daniel M. Greenberger.

In this section we intend to motivate this branch of physics located at the interface of atomic physics, quantum optics and solid state research, and to give a preview of coming attractions. In order to focus on the essential ideas we keep this section brief and postpone more detailed discussions to the later sections.

We start in Sec.~\ref{sec:applications} by mentioning applications of interferometry with cold atoms ranging from tests of the foundations of physics to quantum sensors. We then outline in Sec.~\ref{sec:optical_elements} the realization of optical elements in atom optics such as beam splitters and mirrors leading us in Sec.~\ref{sec:sources_atom_optics} to various sources for our interferometers. Here we emphasize especially the use of an atom chip as a trap and a mirror for laser light opening the avenue towards a quantum tiltmeter and a gravimeter. An outline of our lecture notes in Sec.~\ref{sec:overview} concludes our introduction.

\subsection{Applications of atom interferometry}
\label{sec:applications}

The scope of testing fundamental physics with atom interferometry comprises on the one hand measurements of fundamental constants such as Newton's gravitational constant $G$~\cite{Snadden98PRL,Fixler07Science,Rosi14Nature,Biedermann15PRA} and Sommerfeld's fine-structure constant $\alpha$~\cite{Clade06PRA,Mueller06APB,Bouchendira11PRL,Estey15PRL,Parker191}. Indeed, the result for $\alpha$ obtained with photon-recoil measurements in recent years ~\cite{Bouchendira11PRL} has entered into the determination of the CODATA value. Moreover, a measurement of $\alpha$ has been reported this year~\cite{Parker191} with an accuracy of $2.0\cdot10^{-10}$, which is even more accurate than the best measurements to date, based on measuring the anomalous magnetic moment of the electron~\cite{Hanneke08PRL}. 

On the other hand, testing the pillars of general relativity, for example, the universality of free fall (UFF) resulting from Einstein's equivalence principle~\cite{Damour96CQG}, is of particular interest. The most elementary test of the UFF is to compare the measurements of local gravity with a classical and an atomic~\cite{Peters99Nature,Merlet10Metrologia} gravimeter. 

More elaborate set-ups use two different quantum objects, for instance, two isotopes of the same atomic species, or two different atomic species, and measure their free-fall rate within the same device~\cite{Fray04PRL,Bonnin13PRA,Kuhn14NJP,Tarallo14PRL,Schlippert14PRL,Zhou15PRL}. Future experiments of this kind are expected to catch up to, or even overcome today's best classical tests of the UFF based on Lunar-Laser-Ranging~\cite{Williams12CQG}, torsion balance experiments~\cite{Adelberger09}, or space missions using freely-falling test masses~\cite{Touboul2017PRLMICROSCOPE}. 

In addition, atom interferometers can also test different models in particle physics in the search for unknown forces or dark energy~\cite{Hamilton15Science,Tilburg15PRL,Elder16PRD}. Even more exotic experiments aim for the detection of gravitational waves~\cite{PhysRevLett.110.171102,PhysRevD.78.122002,DIMOPOULOS200937,Hogan2011}, new probes of the foundations of quantum mechanics, such as delayed-choice experiment~\cite{Manning15Nature}, or for the creation of atomic Einstein-Podolsky-Rosen pairs~\cite{Peise15NatComm,Engelsen17PRL}.

Especially in absolute gravimetry, the sensitivity of atomic sensors is competitive with classical devices~\cite{Middlemiss16Nature}. The conventional sensors used for geodesy~\cite{Timmen10Springer} can be categorized as absolute gravimeters, such as the falling-corner cube gravimeters~\cite{Zumberge82Met,Niebauer95Met}, and relative gravimeters like superconducting gravimeters~\cite{Prothero68RSI,Okubo97GRL,Imanishi04Science}, which have a changing bias over time. 

State-of-the-art {\it atomic} gravimeters operate with Raman-type beam splitters and cold atoms, which are either dropped or launched from optical molasses -- a technique invented for Cesium fountain clocks~\cite{Salomon90EPL,Philips91PS}. State-of-the-art laboratory grade examples of these gravimeters~\cite{Hu13PRA,Bidel13APL,Freier16CS,Fang16CS} reach inaccuracies in the low $\upmu$Gal regime. The maturity of this technology has now arrived at a level that commercial products with a specified sensitivity of better than 10\,$\upmu$Gal~\cite{Bodart10APL,muquans,aosense} are available. 

\subsection{Optical elements for atoms}
\label{sec:optical_elements}

The coherent manipulation of matter waves is a central element in every matter wave interferometer~\cite{Cronin09RevModPhys}. Two methods to realize beam splitters and mirrors based on light pulses offer themselves: Raman~\cite{Kasevich91PRL} and Bragg diffraction~\cite{Kozuma99PRL,Torii00PRA}. However, these techniques imply conceptual differences which have to be considered when constructing an atom interferometer aimed at measuring inertial effects~\cite{Antoine03}. 

Indeed, Raman diffraction, where an atomic $\Lambda$-scheme is driven, requires a phase-stable microwave coupling between two hyperfine ground states of an atom usually established by two phase-locked lasers. Working with two different internal states of an atom from an ensemble with a wide velocity distribution has the advantage of velocity filtering with blow-away pulses and state-selective detection~\cite{Kasevich91PRL2,Altin13NJP}. These state-labeling features are described in detail by Christian~Bord\'e~\cite{Borde89PRA}. 

In contrast, Bragg diffraction involves only a single atomic ground state and allows us to construct with a single laser system a pure momentum, or recoil beam splitter. However, due to the transition frequency being in the radio-frequency (RF) range, the detection needs to be spatially resolved, and, in order to distinguish different diffraction orders~\cite{Szigeti12NJP,Altin13NJP}, requires a momentum distribution below recoil.

\subsection{Sources for atom optics}
\label{sec:sources_atom_optics}

Today's generation of atomic inertial sensors typically operates with cold atoms released or launched from an optical molasses. This approach was taken in our simultaneous, dual-species Raman-type interferometer with molasses-cooled \textsuperscript{87}Rb and \textsuperscript{39}K ensembles which measured the E\"otv\"os ratio to $\eta_{\,{\mathrm{Rb},\mathrm{K}}} = (0.3 \pm 5.4) \cdot 10^{-7}$. The velocity distribution and finite size of these sources of atoms limit the efficiency of the beam splitters as well as complicate the analysis of systematic uncertainties.

These limitations can be overcome by the use of atomic ensembles with a typical average momentum well below the recoil of a photon, for example Bose-Einstein condensates (BECs)~\cite{Anderson198,PhysRevLett.75.3969}. The width of a momentum distribution corresponding to a BEC can be further reduced after reaching the regime of ballistic expansion, where all mean field energy is converted to the kinetic energy, by the application of the delta-kick collimation~(DKC) technique~\cite{Muentinga13PRL}. 

Atom-chip technologies offer the possibility to generate a BEC and perform DKC in a fast and reliable way, resulting in miniaturized atomic devices. BECs are very useful for Bragg and double Bragg diffraction \cite{Ahlers16PRL,Abend16PRL} leading to high diffraction efficiencies. Indeed, such beam splitters and mirrors can reach an efficiency of above $95\%$ facilitating interferometry with high contrast.

Furthermore, BECs offer novel methods of coherent manipulation with high fidelity, to realize for example a tiltmeter~\cite{Ahlers16PRL}. A combination of double Bragg diffraction and Bloch oscillations gives rise to a relaunch procedure with an efficiency larger than $75\%$ for the diffraction of atoms in a retro-reflected optical lattice~\cite{Abend16PRL}. The novelty of this method originates from the fact that it relies on a single laser beam, which is also used as a beam splitter, and thus does not lead to an increased complexity of the setup.

We realize a Mach-Zehnder interferometer (MZI) by dropping ensembles directly after release, or accelerating them upwards after a certain time of free fall. The interferometry is performed as in a fountain, such that the total time~$2T$ of the interferometer can be extended. Here~$T$ is the time between the first beam-splitter pulse and the central mirror pulse. 

We utilize an atom chip~\cite{Abend16PRL} for BEC generation and state preparation, including magnetic sub-state transfer, DKC and Stern-Gerlach-type deflection. A special feature of our setup is that the light field, which forms the MZI by Bragg diffraction, is reflected by the atom chip itself. In this way, the chip also serves as an inertial reference inside the vacuum chamber leading to a compact atom-chip gravimeter. 

All atom-optics operations, the interferometry as well as the detection of the output states of the atom interferometer, are integrated into a volume of less than a cube of one centimeter side length. In the fountain mode, the MZI can be extended to have the total interferometer time $2 T = 50$ ms with a large contrast $C = 0 . 8$, which yields an intrinsic sensitivity $\Delta g/g= 1.4\cdot 10^{-7}$. The state preparation comprised of DKC and Stern-Gerlach-type deflection makes an important contribution to this achievement by improving the contrast and reducing the detection noise. An estimation of systematic uncertainties for the current setup and their projection onto a future device prove that it is possible to reach sub-$\upmu$Gal accuracies with a fountain-type geometry.

\subsection{Overview}
\label{sec:overview}
Our lecture notes are organized as follows. In Sec.~\ref{sec:Tools} we introduce the basic tools of atom interferometry such as beam splitters, mirrors, and optical lattices to construct a MZI for atoms. We then turn in Sec.~\ref{sec:EQP} to tests of the equivalence principle. In particular, we present a dual-species atom interferometer for \textup{\Rb} and \textup{\K} to investigate the UFF. Next, we present in Sec.~\ref{sec:BEC_interferometry} interferometers utilizing BECs on an atom chip. We introduce the technique of DKC and present a quantum tiltmeter as well as a gravimeter exploiting this technology. Finally, we conclude in Sec.~\ref{sec:Outlook} by providing an outlook on future devices such as the very long base line atom interferometry (VLBAI) facility and atom interferometers in space.
 
\section{Tools of atom interferometry}
\label{sec:Tools}

An atom interferometer requires the realization of a beam splitter and a mirror for atom waves. The underlying processes have to be coherent and phase-stable in order to establish an interference pattern.
Several options for such elements exist, and we analyze prominent ones exploited in current experiments below. In particular, in Sec. \ref{sec:beam_splitters_mirrors} we discuss beam splitters and mirrors based on Bragg and Raman diffraction. Next, in Sec. \ref{sec:Optical_lattice} we focus on the manipulation and accelerations of atoms by optical lattices. Concluding, we introduce in Sec.~\ref{sec:Mach_Zehnder} a common interferometer geometry based on beam splitters and mirrors.

\subsection{Beam splitters and mirrors}
\label{sec:beam_splitters_mirrors}

Beam splitters and mirrors for matter waves can be realized with the help of mechanical gratings~\cite{PhysRevLett.66.2693,PhysRevLett.66.2689}, or electromagnetic waves~\cite{Kasevich91PRL,Kasevich92APB,PhysRevLett.67.177,Rasel95PRL,Berman}. While single-photon electric or magnetic dipole transitions can implement a coherent electromagnetic coupling, the assessment in Sec.~\ref{sec:Rabi_oscillations} focuses on stimulated two-photon transitions. For this purpose, we first introduce the Rabi model which describes the atom-light interaction in an effective two-level system before we proceed to two-photon transitions in a three-level system. Here the absorption of a photon from a field with frequency $\omega_1$ is followed by stimulated emission into a field with frequency $\omega_2$, and {\it vice versa}. 

Next, we consider in Sec.~\ref{sec:Bragg_Raman_diffraction} the momentum transfer due to the atom-light interaction which is at the very heart of the sensitivity of atom interferometers to inertial forces. We also discuss two standard approaches towards beam splitters. Raman diffraction is a widely used technique designed for atoms which have been laser-cooled in optical molasses without the application of additional cooling steps. Bragg diffraction is a powerful tool for delta-kick collimated Bose-Einstein condensates, since the velocity dispersion of the ensemble is of major relevance for the manipulation efficiency. 

In Sec.~\ref{sec:Multi-photon_Bragg} we then concentrate on the generalization of a multi-photon coupling utilizing Bragg diffraction, and conclude in Sec.~\ref{sec:FiniteSize}, where we take into account the effects of the finite size of the atom cloud and the laser beam on the atom-light interaction.

\subsubsection{Rabi oscillations and two-photon coupling}
\label{sec:Rabi_oscillations}
We consider the atom as an effective two-level system consisting of the internal states $\ket{\mathrm{g}}$ and $\ket{\mathrm{e}}$ with energies $\hbar \omega_{\mathrm{g}}$ and $\hbar \omega_{\mathrm{e}}$, respectively, and a dipole moment $\vec{d}$. The time evolution of the state populations under the influence of a resonant ($\omega_0=\omega_{\mathrm{eg}}\equiv \omega_\mathrm{e}-\omega_\mathrm{g}$) electromagnetic field~$\vec{E}\equiv\vec{E}_0 \textnormal{cos}(\omega_{0}\tau + \phi)$ at time~$\tau$ with frequency~$\omega_{0}$ and phase~$\phi$ is determined by the Rabi frequency
\begin{equation} 
 \Oeg \equiv \frac{\bra{\mathrm{e}} \vec{d} \cdot \vec{E}_0 \ket{\mathrm{g}} } {\hbar}= \Gamma \sqrt{\frac{I}{2\Isat}}\,, 
 \label{eq:rabi}
\end{equation}
expressed in terms of the intensity~$I$ of the light field with the saturation intensity~$\Isat$, and the natural linewidth~$\Gamma$ of the transition. Indeed,~$\Oeg$ is assumed to be constant and is a measure of the coupling strength between the atom modeled by the two atomic states~$\ket{\mathrm{g}}$ and~$\ket{\mathrm{e}}$, and the electromagnetic field $\textbf{E}$. 

Off-resonant driving with a non-vanishing detuning~$\delta\equiv\oeg-\omega_0$ is taken into account in the effective Rabi frequency
\begin{equation} \label{eq:rabieff} 
\Oeff \equiv \sqrt{|\Oeg|^2 +\delta ^2} \,, 
\end{equation} 
which always leads to a faster oscillation of the probability
\begin{equation} \label{eq:probrabieff}
 P_{\mathrm{e}} (\tau,\delta,\Oeg) = \frac{1}{2}\left(\frac{\Oeg}{\Oeff}\right)^2[1 - \mathrm{cos}(\Oeff \tau)]
\end{equation}
to find after an interaction time~$\tau$ the atom in the excited state  $\ket{\mathrm{e}}$ if the atom is initially prepared in the ground state $\ket{\mathrm{g}}$.

The reduced amplitude of the oscillation is determined by the ratio $\Oeg/\Oeff$ of the resonant and the effective Rabi frequency.  For a vanishing detuning, that is $\delta=0$, the amplitude of $P_{\mathrm{e}}$ is unity, whereas for a large detuning, $|\delta|\gg |\Oeg|$, the amplitude of $P_{\mathrm{e}}$ tends towards zero.

Rabi oscillations can be driven efficiently only for long-lived states, that is for states in which~$\Oeff$ is large compared to the inverse lifetime of the working states $\ket{\mathrm{g}}$ and $\ket{\mathrm{e}}$. A common method to avoid decays from~$\ket{\mathrm{e}}$ to ~$\ket{\mathrm{g}}$ is to choose these states such that the selection rules forbid a single-photon transition between them. The two states could then be coupled via a two-photon transition which requires an intermediate state~$\ket{\mathrm{i}}$. 

\begin{figure}[h]
\centering
\includegraphics[width=0.5\linewidth]{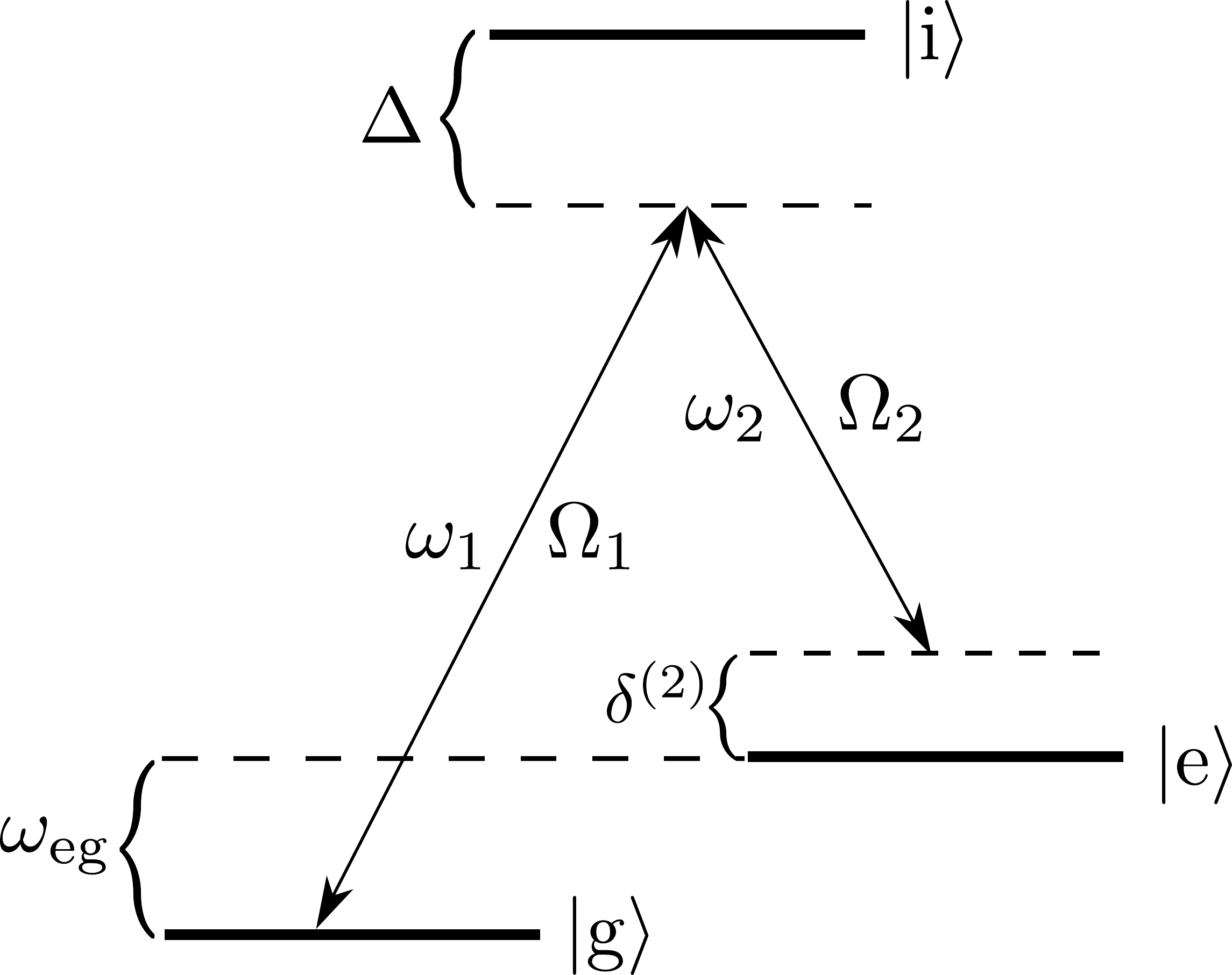}
\caption{Three-level atom interacting with two light fields. A two-photon coupling between the atomic states $\ket{\mathrm{g}}$ and $\ket{\mathrm{e}}$ is established by two electromagnetic fields of frequencies $\omega_1$ and $\omega_2$, with $\Omega_1$ and $\Omega_2$ being the Rabi frequencies of the corresponding one-photon transitions. Here $\Delta$ is the common detuning of the two-photon transition from the intermediate state $\ket{\mathrm{i}}$, and $\delta^{(2)}$ is the two-photon detuning.}
\label{fig:Lambda_Scheme}
\end{figure}

As depicted in Fig.~\ref{fig:Lambda_Scheme}, light fields with frequencies $\omega_1$ and $\omega_2$ induce non-resonant transitions, where the transitions $\ket{\mathrm{g}}\longleftrightarrow \ket{\mathrm{i}}$ and $\ket{\mathrm{i}}\longleftrightarrow \ket{\mathrm{e}}$ are detuned by $\Delta$ and $\Delta+\delta^{(2)}$, respectively. Consequently, the frequency difference $\delta\omega\equiv\omega_1-\omega_2$ is equal to the frequency difference $\oeg$ between the working states plus the two-photon detuning $\delta^{(2)}$, that is $\delta\omega\equiv\oeg+\delta^{(2)}$. Here and in the following, the superscript $^{(2)}$ indicates a two-photon process.

The intermediate state can now be short-lived itself, since it is only virtually populated, but enables sufficient coupling via simultaneous stimulated absorption and emission. The resulting two-photon Rabi frequency $\Omega_{12}$, which drives the transition $\ket{\mathrm{g}}\longleftrightarrow \ket{\mathrm{e}}$ in the case $\Delta \gg \Omega_j$ with~$j=1,2$, is governed by the product of both Rabi frequencies $\Omega_1$ and $\Omega_2$ as well as the common detuning~$\Delta$ of the two-photon transition to the intermediate state~$\ket{\mathrm{i}}$, that is
\begin{equation} \label{eq:twophotonrabi} 
\Omega_{12} \equiv \frac{\Omega _1 ^* \Omega_2}{2\Delta} = \frac{\Gamma_1 \Gamma_2}{4 \Delta} \sqrt{\frac{I_1}{I_{\textrm{sat},1}}\frac{I_2}{ I_{\textrm{sat},2}}}\,.
\end{equation}
Here~$I_j$ and~$I_{\textrm{sat},j}$ denote the intensity and the saturation intensity of the corresponding light beam, and~$\Gamma_j$ is the natural linewidth of the corresponding transition.

In this case we return to an effective two-level system where the probability 
\begin{equation} \label{eq:probrabieff_12}
 P_{\mathrm{e}} \left(\tau,\delta^{(2)},\Omega_{12}\right) = \frac{1}{2}\left(\frac{\Omega_{12}}{\Oeff^{(2)}}\right)^2\left[1 - \mathrm{cos}\left(\Oeff^{(2)} \tau\right)\right]
\end{equation}
to find the atom in the excited state $\ket{\mathrm{e}}$ now depends on the two-photon Rabi frequency $\Omega_{12}$, and the corresponding effective Rabi frequency
\begin{equation} \label{eq:rabieff_12}
\Oeff^{(2)}=\sqrt{\left|\Omega_{12}\right|^2+\left(\delta^{(2)}\right)^2}\,,
\end{equation}
determined by the two-photon detuning $\delta^{(2)}$.

A fundamental loss mechanism of the coherent dynamics is spontaneous emission which is fortunately suppressed due to the fact that the two-photon transition is off-resonant by the detuning~$\Delta$ relative to the intermediate state~$\ket{\mathrm{i}}$. The rate~$R_{\mathrm{sp}}$ of residual spontaneous decay then reads
\begin{equation}\label{eq:spontan} 
R_{\mathrm{sp}} \equiv\frac{\sqrt{\Gamma_1\Gamma_2}}{2\Delta}\left|\Omega_{12}\right|=\frac{(\Gamma_1\Gamma_2)^{\frac{3}{2}}}{8\Delta^2}
\sqrt{\frac{I_1}{I_{\mathrm{sat},1}}\frac{I_2}{I_{\mathrm{sat},2}}}\,.
\end{equation}
For sufficiently large detuning it is possible to suppress spontaneous emission almost completely. 

Moreover, the presence of an off-resonant light field has an influence on the atomic energy structure. Indeed, the one-photon ac-Stark shift causes an energy shift
\begin{equation}\label{eq:acstark}
\delta E_j^{\mathrm{ac}}=-\frac{\hbar \left|\Omega_j\right|^2}{4\Delta_j}
\end{equation}
of the undisturbed atomic states~$\ket{\mathrm{g}} (j=1)$ and~$\ket{\mathrm{e}} (j=2)$, determined by the detuning~$\Delta_j$, and the Rabi frequency~$\Omega_j$ of the corresponding transition~\cite{Foot} with $\Delta_1\equiv \Delta$ and $\Delta_2\equiv\Delta+\delta^{(2)}$.

Furthermore, for high-precision measurements such as the ones discussed in these lectures also the two-photon light shift has to be considered, which depends on the details of the internal atomic structure, as well as on the polarization of the light fields~\cite{Berg14,Schlippert14,Giese16PRA}. 

Finally, we consider two special cases of the Rabi dynamics given by Eq.~\eqref{eq:probrabieff_12}, namely \quot{$\pi/2$}- and \quot{$\pi$}-pulses, which are determined by their enclosed pulse areas. For fixed laser intensities~$I_1$ and~$I_2$, we define these pulses by their specific interaction times~$\tau_{\pi/2}\equiv\pi/\left(2\,\Oeff^{(2)}\right)$ and~$\tau_{\pi}\equiv\pi/\Oeff ^{(2)}$, leading us to the probabilities
\begin{equation}
P_{\mathrm{e}}\left(\tau_{\pi/2},\delta^{(2)},\Omega_{12}\right)=\frac{1}{2}\left(\frac{\Omega_{12}}{\Oeff^{(2)}}\right)^2
\end{equation}
and
\begin{equation}
P_{\mathrm{e}}\left(\tau_{\pi},\delta^{(2)},\Omega_{12}\right)=\left(\frac{\Omega_{12}}{\Oeff^{(2)}}\right)^2\,,
\end{equation}
where we have made use of  Eq.~\eqref{eq:probrabieff_12}.

In the ideal case with $\delta^{(2)}=0$ a $\pi/2$-pulse creates an equally weighted superposition of $\ket{\mathrm{g}}$ and $\ket{\mathrm{e}}$ when starting in one of the two working states. In contrast, a $\pi$-pulse inverts the states $\ket{\mathrm{g}}$ and $\ket{\mathrm{e}}$. Due to their functions in an interferometer, these pulses are called \quot{beam splitter} and \quot{mirror} for atoms, in complete analogy to their counterparts in optics for light beams.

\subsubsection{Bragg and Raman diffraction}
\label{sec:Bragg_Raman_diffraction}
So far we have only discussed the dynamics of {\it internal} atomic states induced by the atom-light interaction. However, the use of an atom interferometer for inertial sensing requires sensitivity to {\it external} degrees of freedom, in particular, the atomic center-of-mass motion relative to a reference frame. We satisfy this requirement when we recall that during the atom-light interaction the electromagnetic field does not only transfer energy, but also momentum to the atoms. 

We now consider a two-photon process induced by two light fields with the wave vectors $\vec{k}_1$ and $\vec{k}_2$. In the case of counter-propagating  fields, that is~$\vec{k}\equiv\vec{k}_1\approx-\vec{k}_2$, the momentum transfer between atom and field is maximal and approximately $2\hbar \vec{k}$. For co-propagating beams, that is~$\vec{k}_1\approx\vec{k}_2$, the momentum transfer is minimal and almost zero. 

Furthermore, due to the fact that the dispersion relation of a free particle is parabolic, a non-zero momentum~$\vec{p}_0$ of the atom and the resulting frequency shift have to be taken into account. Indeed, any offset~$\vec{p}_0$ results in a Doppler shift
\begin{equation} \label{eq:dopplerfrequency} 
\omega_{\mathrm{D}} \equiv \frac{\vec{p}_0\cdot \keffvc}{m}
\end{equation}
of the transition frequencies due to the motion of the atoms of mass~$m$ relative to the light fields. It vanishes only for atoms at rest. 

These considerations also lead us to the definition
\begin{equation} \label{eq:recoildoppler} 
\orec \equiv \frac{\hbar \left|\keffvc\right|^2}{2m}
\end{equation}
of the recoil frequency associated with the light fields. Here, and in Eq.~\eqref{eq:dopplerfrequency} we have introduced the notation~$\keffvc\equiv\vec{k}_1-\vec{k}_2$ to identify the effective momentum transfer~$\hbar \keffvc$ during a two-photon process.

In the general case of an $n$\textsuperscript{th}-order transition and counter-propagating light fields with $\vec{k}\equiv\vec{k}_1\approx-\vec{k}_2$, the total momentum transfer 
\begin{equation}\label{eq:keff}
n \hbar \keffvc\equiv n \hbar\left(\vec{k}_{\text{1}}-\vec{k}_{\text{2}}\right)\approx 2n\hbar \vec{k}
\end{equation}
is the sum of the momenta transferred by $n$ photon pairs and can achieve large values. 

In a quantum mechanical treatment of the atomic center-of-mass motion, an $n$\textsuperscript{th}-order two-photon transition couples the momentum eigenstates $\ket{\vec{p}_0}$, corresponding to the momentum~$\vec{p}_0$ before the interaction, and $\ket{\vec{p}_n}$, representing the momentum $\vec{p}_n\equiv\vec{p}_0+n\hbar\keffvc$ after the interaction. Based on the considerations of Sec.~\ref{sec:Rabi_oscillations}, we also have to include a coupling to the internal states. However, since a change of a momentum eigenstate does not necessarily require a change of an internal atomic state, different types of diffraction are possible~\cite{Borde89PRA}. 

We call an atomic scattering process a \quot{Raman}-type diffraction if the atom-light interaction couples two different internal states of the atom, whereas we call it a \quot{Bragg}-type diffraction if the internal state is unchanged. More sophisticated schemes employ double Raman diffraction \cite{Leveque09PRL,Malossi10PRA}, or double Bragg diffraction \cite{Giese13PRA,Ahlers16PRL}. They lead to a larger momentum transfer, and hence provide us with an increased sensitivity as well as the elimination of certain noise sources due to the symmetric structure of the diffraction process.

\subsubsection{Multi-photon coupling by Bragg diffraction}
\label{sec:Multi-photon_Bragg}
Bragg diffraction of a matter wave is defined in complete analogy to the diffraction of an electromagnetic field by a crystal~\cite{Bragg12Nature,Friedrich13AP}. Here the roles of light and matter are interchanged. 

Indeed, when an atomic beam or ensemble is diffracted from two counter-propagating light fields of the frequencies $\omega_1$ and $\omega_2$ the Bragg condition reads
\begin{equation}\label{eq:braggcondition} 
\delta E_{\mathrm{kin}} = n \hbar\delta\omega\equiv n\hbar (\omega_1-\omega_2)\,,
\end{equation}
where 
\begin{equation}\label{eq:delta_Ekin}
\delta E_{\mathrm{kin}}\equiv \frac{\left(\vec{p}_0+n\hbar \vec{k}_\mathrm{eff}\right)^2}{2m}-\frac{\vec{p}_0^2}{2m}
\end{equation}
is the change of the kinetic energy of the atom associated with its change in momentum.

\begin{figure}[h]
\includegraphics[width=\linewidth]{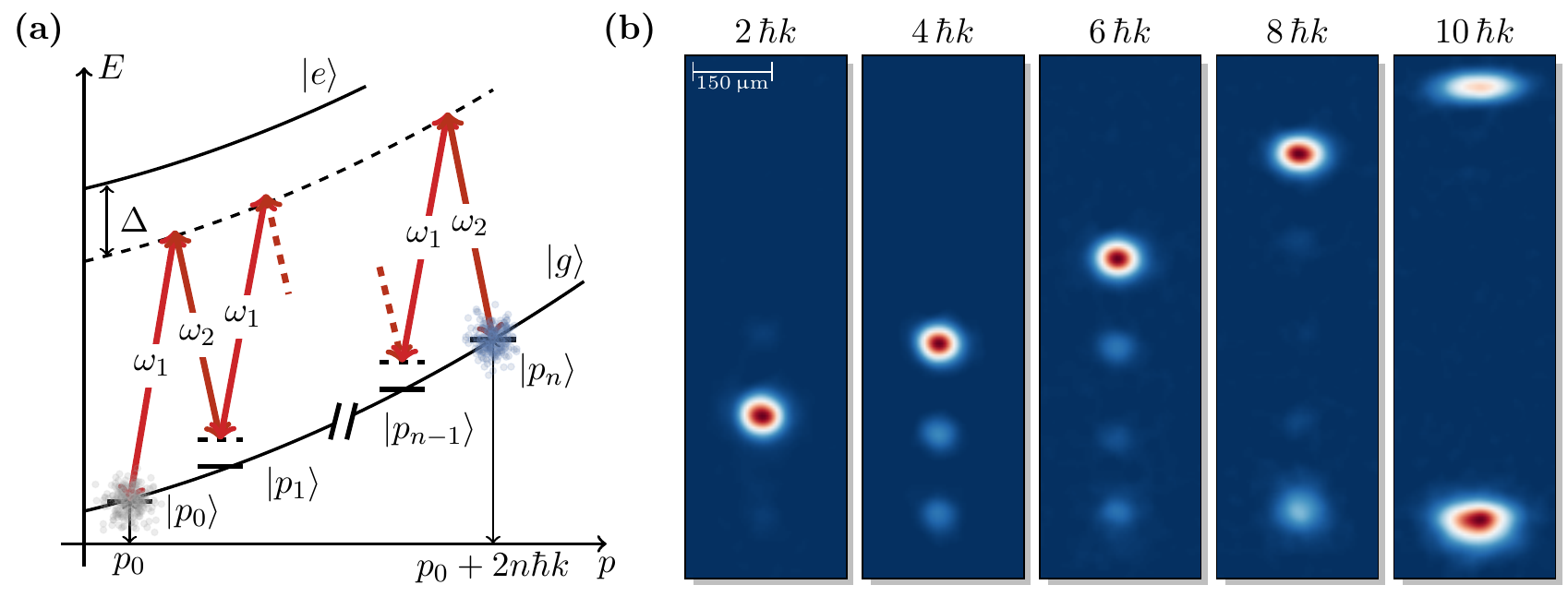}
\caption{Momentum transfer of~$2n\hbar k$ in $n$\textsuperscript{th}-order Bragg diffraction represented by the level scheme~(a) and density plots~(b) measured experimentally for $n=1,2,3,4$ and $5$. This figure is an adaptation of Figs.~4.7 and 5.12 in Ref.~\cite{Abend17}.}
\label{fig:higherorderbragg}
\end{figure}

For the case of first-order diffraction, $n=1$, Eqs.~\eqref{eq:braggcondition} and~\eqref{eq:delta_Ekin} give a very intuitive picture of the scattering process. An atom scatters two photons with momenta~$\hbar\vec{k}_1$ and~$\hbar\vec{k}_2$ from two traveling light waves if their energy difference~$\hbar\delta\omega$ matches the energy~$\delta E_{\mathrm{kin}}$ an atom has to absorb to climb the kinetic energy parabola, depicted in Fig.~\ref{fig:higherorderbragg}(a). 

When we use Eqs.~\eqref{eq:braggcondition} and~\eqref{eq:delta_Ekin} and the definitions Eqs.~\eqref{eq:dopplerfrequency} and~\eqref{eq:recoildoppler} of the Doppler shift~$\omega_{\mathrm{D}}$ and the photon recoil~$\omega_{\mathrm{rec}}$, we arrive at the condition
\begin{equation}\label{eq:braggfrequency} 
\omega_1-\omega_2 = n\, \omega_{\mathrm{rec}} + \omega_{\mathrm{D}}\,,
\end{equation}
for the frequency difference which drives the diffraction process of the~$n$-th order. 

The case of scattering more than one photon pair at a time, $n>1$ leads to the population of higher-order momentum states~$\ket{\vec{p}_n}$ with $\vec{p}_n =\vec{p}_0+n\hbar\vec{k}_{\mathrm{eff}}$, since ideally every momentum state in between is not on resonance and should not be populated, as shown in Fig.~\ref{fig:higherorderbragg}(a).

In Fig.~\ref{fig:higherorderbragg}(b) we present density plots for multi-photon Bragg diffraction with~$n=1,2,3,4$ and $5$, and note that the simultaneous scattering of $n$ pairs of photons has been realized experimentally up to $n=12$. This achievement allows us to construct a beam splitter~\cite{Mueller08PRL} with a momentum transfer of~$24\,\hbar k$, where $k=\left|\vec{k}\right|$. For \textsuperscript{87}Rb the spacing between subsequent Bragg orders is only 15\,kHz, which leads to populating multiple orders. 

For the $n$\textsuperscript{th}-order Bragg transition the calculation of the transition probability $P_n$ requires considerations~\cite{Mueller08PRA} that go beyond the two-state assumption. The product of the Rabi frequency~$\Oeff$ governed by the laser intensity~$I$, and the duration~$\tau$ of the atom-light interaction are the major ingredients. Indeed, the product of these parameters determines if a clean Rabi oscillation into a single momentum state is possible, or if multiple states are populated.

In the Bragg regime where a single order is dominantly populated, the generalized transition probability
\begin{equation} 
P_{n}(\tau)\equiv\textnormal{sin}^2\left[\frac{1}{2} \int_0^\tau \diff\tau'\Omega_n(\tau') \right]
\end{equation}
from the initial momentum state $\ket{\vec{p}_0}$ to the $n$\textsuperscript{th}-order momentum state $\ket{\vec{p}_n}$ is determined by a new effective Rabi frequency $\Omega_n$, as discussed in detail in Ref.~\cite{Mueller08PRA}.

In order to perform an $n$\textsuperscript{th}-order transition an increase in the laser power is required. An approximate solution obtained in Ref.~\cite{Mueller08PRA} presents conditions for a so-called \quot{quasi-Bragg} regime, in which the probability to populate a single higher-order state is significantly larger compared to the one for a population of all other orders by applying short and intense pulses. 

In order to achieve a large momentum transfer without the increase of laser power, one can use sequential pulses. Indeed, the same momentum transfer as in the case of a single $n$\textsuperscript{th}-order transition may be achieved at the cost of a larger total time of the beam splitting process, and a more complex waveform.
For a sequence consisting of $n_p$ sequential pulses the resonance condition to drive the $n_s$-th sequential transition with an $n$\textsuperscript{th}-order Bragg pulse reads
\begin{equation}\label{eq:seqtrans}
\omega_1-\omega_2=(2n_s-1)n\, \orec+\omega_{\mathrm{D}}\,.
\end{equation}
with $n_s=1,\ldots,n_p$. We emphasize that the frequency difference $\omega_1-\omega_2$ has to be adjusted for each sequential transition. In principle, there is no restriction on combining any number $n_p$ of sequential pulses with any achievable Bragg order $n$, to obtain a particular transfer efficiency. For example, a sequence of beam splitters transferring $6\, \hbar k$ each, leads to a total splitting~\cite{Chiow11PRL} of $102\, \hbar k$, or a sequence of first-order transitions results~\cite{Kovachy15Nature} in $90\, \hbar k$. 

\subsubsection{Influence of atom cloud and beam size}
\label{sec:FiniteSize}
In the single-atom picture, or in the case of a highly monochromatic ensemble, one can always find for fixed laser powers~$I_1$, $I_2$ and a common detuning~$\Delta$, an interaction time~$\tau$, for which the amplitude of the transition probability given by Eqs. \eqref{eq:probrabieff_12} and \eqref{eq:rabieff_12} is unity. Here we assume that the two-photon detuning~$\delta^{(2)}$ vanishes. In this case the beam-splitter efficiency is only limited by the loss of atoms due to spontaneous decay with the rate~$R_{\mathrm{sp}}$, expressed by Eq. \eqref{eq:spontan}.

However, a non-zero temperature~$T_\mathrm{a}$ of the atoms, and therefore a spread~$\sigma_v$ in the velocity~$\vec{v}$ of the atomic ensemble needs to be taken into account, as it induces a broadening of the transition frequency due to the Doppler shift $\vec{k}_{\mathrm{eff}}\cdot \vec{v}$, Eq. \eqref{eq:dopplerfrequency}. Even for BECs, the beam-splitter efficiency may change drastically, when we employ for example higher-order Bragg diffraction~\cite{Szigeti12NJP}. 

We estimate the influence of such a velocity distribution by the use of a Gaussian distribution 
\begin{equation}\label{eq:3Dgaussian}
f_{3\mathrm{D}}(\vec{v})\equiv\frac{1}{(2\pi)^{3/2}\sigma_v^3}\mathrm{exp}\left[- \frac{(\vec{v}-\vec{v}_0)^2}{2\sigma_v^2}\right]
\end{equation}
of velocities $\vec{v}$ across the atomic ensemble which is isotropic in all three spatial dimensions and has a width
\begin{equation}
\sigma_v \equiv \sqrt{\frac{k_{\mathrm{B}} T_\mathrm{a}}{m}}
\end{equation}
determined by the temperature~$T_\mathrm{a}$ and the Boltzmann constant~$k_{\mathrm{B}}$. Here $\vec{v}_0$ is an arbitrary offset velocity.

The total probability~$P_{\mathrm{e}}$ to find the atom in the excited state $\ket{\mathrm{e}}$ then reads
\begin{equation}\label{eq:velocitydependence}
P_{\mathrm{e}}(\tau)\equiv \iiint \diff^3 v\, f_{3\mathrm{D}}(\vec{v})P_{\mathrm{e}}\left[\tau,\delta^{(2)}(\vec{v}),\Omega_{12}\right] \,,
\end{equation}
where $P_{\mathrm{e}}\left[\tau,\delta^{(2)}(\vec{v}),\Omega_{12}\right]$ is the excitation probability given by Eq.~\eqref{eq:probrabieff_12} for the atomic velocity~$\vec{v}$, and $\textrm{d}^3 v$ is the three-dimensional volume element in velocity space. 

In order to evaluate the integral over~$\vec{v}$, we recall that according to Eq.~\eqref{eq:probrabieff_12} the probability $P_{\mathrm{e}}\left[\tau,\delta^{(2)}(\vec{v}),\Omega_{12}\right]$ is determined by the two-photon detuning~$\delta^{(2)}$ which enters into the effective Rabi frequency $\Oeff^{(2)}$ given by Eq.~\eqref{eq:rabieff_12}. We assume here that the two-photon detuning~$\delta^{(2)}\equiv \vec{k}_{\mathrm{eff}}\cdot \vec{v}$ is solely induced by the Doppler shift, and hence depends only on the projection of the velocity~$\vec{v}$ onto the wave vector~$\keffvc$. 

Thus, the three-dimensional integral, Eq. \eqref{eq:velocitydependence}, reduces to a one-dimensional integral over the $x$-component~$v_x$ of the velocity $\vec{v}$ if we align the $x$-axis with the direction of~$\keffvc$, giving rise to
\begin{equation} \label{eq:1Dprop}
P_{\mathrm{e}}(\tau) = \int\diff v_x\, f_{1\mathrm{D}}(v_x)P_{\mathrm{e}}\left[\tau,\delta^{(2)}(v_x),\Omega_{12} \right]
\end{equation}
with the one-dimensional Gaussian distribution
\begin{equation}
f_{1\mathrm{D}}(v_x)\equiv\frac{1}{(2\pi)^{1/2}\sigma_v}\mathrm{exp}\left[- \frac{(v_x-v_{x,0})^2}{2\sigma_v^2}\right]\,.
\end{equation}
We emphasize that the $x$-component~$v_{x,0}$ of the offset velocity~$\vec{v}_0$ can be compensated by an adjustment of the frequency difference $\delta\omega$ of the two light fields.

Unfortunately, the velocity spread is not the only effect we need to account for. Indeed, the atomic ensemble has also a finite size and interacts with two laser beams having for example Gaussian intensity profiles
\begin{equation}
I_j(y,z)\equiv I_{0,j}\exp\left[-\frac{2(y^2+z^2)}{w_{j}^2}\right]
\end{equation}
in the $y-z$-plane, where~$I_{0,j}$ is the amplitude and~$w_{j}$ is the radius of the beam for $j=1,2$. 

The finite size of the laser beams causes a spatially dependence~$\Omega_{12}=\Omega_{12}(\vec{r})$ of the Rabi frequency due to the dependence~$I_j=I_j(\vec{r})$ of the intensity on the position $\vec{r}$ of the atom within the two beams. As a result, for two collimated laser beams aligned along the $x$-direction, the excitation probability
\begin{equation}
P_{\mathrm{e}}\left[\tau,\delta^{(2)},\Omega_{12}(\vec{r})\right]=P_{\mathrm{e}}\left[\tau,\delta^{(2)},\Omega_{12}(y,z)\right]
\end{equation}
only depends on the coordinates $y$ and $z$ perpendicular to~$\keffvc$.

Moreover, we model the spatial transverse distribution of the atomic ensemble by a Gaussian
\begin{equation}\label{eq:spatialdistribution}
s_{2\textrm{D}}(y,z)\equiv\frac{1}{2\pi\sigma_y^2}\mathrm{exp}\left(-\frac{y^2+z^2}{2\sigma_y^2}\right)
\end{equation}
of width~$\sigma_y=\sigma_z$ centered at the maximum intensity of the beams ($y=z=0$) in the $y-z$-plane. 

When we combine Eqs.~\eqref{eq:1Dprop} and \eqref{eq:spatialdistribution}, the total probability in three dimensions reads
\begin{equation}\label{eq:3Dprop}
P_{\textrm{e}}(\tau) \equiv \iint \diff y\diff z \int \diff v_x \,s_{2\mathrm{D}}(y,z) f_{1\mathrm{D}}(v_x) P_{\mathrm{e}}\left[\tau,\delta^{(2)}(v_x),\Omega_{12}(y,z)\right],
\end{equation}
and is determined by the widths~$\sigma_v$, $\sigma_y$, the laser beam radii~$w_{j}$ as well as the Rabi frequency~$\Omega_{\textrm{12}}$ depending on the maximum laser intensity~$I_{0,j}$. 

Usually, the velocity distribution~$f_{1\mathrm{D}}$ can be assumed to be time-independent if no external force is acting. However, the spatial distribution~$s_{2\textrm{D}}$ is always a function of time, since the cloud spreads due to the non-vanishing width~$\sigma_v$ of the velocity distribution. Hence, finding the three-dimensional probability~$P_{\mathrm{e}}$ is more complicated if these assumptions are not valid. Even in the case of perfect monochromatic light fields, the efficiency of the coherent processes is fundamentally limited by the finite size and velocity of the cloud of atoms~\cite{Szigeti12NJP}.

\subsection{Optical lattices}
\label{sec:Optical_lattice}

In the preceding sections we have discussed the possibility of changing the atomic momentum with the help of the atom-light interaction leading to Bragg and Raman diffraction. However, there exists also the option of a {\it sequential} momentum transfer in an optical lattice. In particular, a large effect occurs due to Bloch oscillations in an accelerated optical lattice \cite{Dahan96PRL,Peik97PRA,Wilkinson96PRL,Raizen97PT}.

\subsubsection{Bloch theorem}
In one space dimension we can obtain an optical lattice by retroreflecting a light field propagating in the $x$-direction and having the wave vector $\vec{k}$, from a mirror. This process leads to the formation of a standing light wave, and hence, to an effective periodic potential
\begin{equation}\label{eq:standingwavepotential} V(x)\equiv  4 V_{\mathrm{dip}}\ \mathrm{sin}^2\left(k x\right) = \frac{1}{2} V_0 \left[1-\cos\left(2 k  x\right)\right]
\end{equation}
for atoms with the amplitude~$V_0=4 V_{\mathrm{dip}}$, where $V_{\mathrm{dip}}$ is the magnitude of the atom-light interaction~\cite{Grimm00AAMOP}. The factor of four results from the amplification of the electric field~$\vec{E}$ by a factor of two due to the retro-reflection, 
and the quadratic scaling of the light field intensity~$I \propto |\vec{E}|^2$. 

The potential given by Eq.~\eqref{eq:standingwavepotential} is periodic with the period~$d\equiv\pi/k$ given in units of the wave number~$k\equiv\left|\vec{k}\right|\equiv 2\pi/\lambda$, and the amplitude
\begin{equation}\label{eq:latticedepth} 
V_0\equiv \frac{\hbar \Gamma^2}{2\Delta}\frac{I}{I_{\textrm{sat}}} 
\end{equation}
of the potential~$V$ can be expressed in terms of the intensity~$I$, the saturation intensity~$I_{\textrm{sat}}$, the natural linewidth~$\Gamma$, and the detuning~$\Delta$. Here we have used Eq.~\eqref{eq:rabi} and Eq.~\eqref{eq:acstark} for the ac-Stark shift.

According to the Bloch theorem \cite{kittel05}, the wave function
\begin{equation}\label{eq:wavefunc_blochwave}
\psi_{\ell,q}(x)\equiv\textrm{e}^{\textrm{i} q x} u_{\ell,q}(x)
\end{equation}
of an atom in a periodic potential $V$, Eq.~\eqref{eq:standingwavepotential}, is the product of a plane wave~$\textrm{e}^{\textrm{i}qx}$ with the quasi-momentum~$\hbar q$, and an amplitude~$u_{\ell,q}(x)\equiv\bra{x}u_{\ell,q}\rangle$ having the same period~$d$ as the original potential~$V$. Here~$\ell$ denotes the discrete band index. 

The Bloch state $\ket{u_{\ell,q}}$ obeys the Schr\"odinger equation
\begin{equation} \label{eq:blochwave}
\left[\frac{(\hat{p}+\hbar q)^2}{2m} +V(\hat{x})\right]\ket{u_{\ell,q}} = E_\ell(q)\ket{u_{\ell,q}},
\end{equation}
where the corresponding quasi-energy
\begin{equation}\label{eq:quasi_energy}
E_\ell(q)=E_\ell\left(q+\frac{2\pi}{d}\right)
\end{equation}
has a period $2\pi/d=2k$ as a function of $q$. Therefore, following the convention of solid-state physics, the quasi-momentum $\hbar q$ can be restricted to the interval~$(-\pi\hbar/d,+\pi\hbar/d]=(-\hbar k,+\hbar k]$, that is the first Brillouin zone~\cite{kittel05}.

In order to work in this interval, it is natural to consider atomic ensembles with a narrow momentum distribution, that is $m\sigma_v\ll\hbar k$. As an example, both a BEC~\cite{Denschlag02JPB}, and a distribution of cold atoms prepared by a velocity filter in one dimension~\cite{Dahan96PRL,Peik97PRA} fulfill this condition.

\subsubsection{Bloch oscillations}
The Bloch states~$\ket{u_{\ell,q}}$ following from Eq.~\eqref{eq:blochwave} and describing an atom in an optical lattice are stationary ones. Therefore, dynamics only occurs if an additional force~$F$ is acting, which can be either an external force, such as gravity, or an acceleration of the lattice itself. 

When an atom in an optical lattice is suddenly exposed to a spatially uniform force~$F$, the Bloch states~$\ket{u_{\ell,q}}$ are no longer eigenstates~\cite{Dahan96PRL,Peik97PRA} of the new Hamiltonian
\begin{equation} \label{eq:forceschroedinger}
\hat{H}_{F} \equiv \frac{\hat{p}^2}{2m} + V(\hat{x}) -F\hat{x}\,. 
\end{equation}

Indeed, for a small value of $F$, such that there are no inter-band transitions and the adiabatic assumption is valid, the wave function $\psi_{\ell,q(0)}$, Eq. \eqref{eq:wavefunc_blochwave}, evolves into 
\begin{equation} 
\psi_{\ell,q(\tau)}(x)= \exp\left\{-\frac{\textrm{i}}{\hbar}\int_0^\tau E_\ell\left[q(t)\right]\diff t\right\} \textrm{e}^{\textrm{i}q(\tau)x}  u_{\ell,q(\tau)}(x)\,,
\end{equation}
and, apart from a phase factor and a linear shift of the quasi-momentum
\begin{equation}\label{eq: time_dep_quasi_momentum}
\hbar q(\tau)\equiv\hbar q(0)+F\tau\,,
\end{equation}
preserves its original form.

Because~$\hbar q$ changes linearly in time $\tau$, the wave function~$\psi_{\ell,q(\tau)}$ has a temporal periodicity given by the Bloch period
\begin{equation}\label{eq:Bloch_period}
\tau_{\mathrm{Bloch}}\equiv\frac{2\pi\hbar}{|F|d}\,,
\end{equation}
which is the time after which the change $F\tau_{\mathrm{Bloch}}$ of the quasi-momentum is equal to the width $2\pi \hbar/d$ of the first Brillouin zone, displayed in Fig.~\ref{fig:blochoscillations}(a). 

Moreover, since the quasi-energy $E_\ell=E_\ell(q)$, Eq.~\eqref{eq:quasi_energy}, is a periodic function with period~$2\pi/d$, the velocity
\begin{equation}
\langle v_\ell \rangle\left(q \right)\equiv\frac{1}{\hbar}\frac{\textrm{d} E_\ell(q)}{\textrm{d} q}
\end{equation}
of an atom in the Bloch wave $u_{\ell,q(\tau)}$ has the same periodicity.

Furthermore, according to Eq.~\eqref{eq: time_dep_quasi_momentum} the quasi-momentum~$\hbar q=\hbar q(\tau)$ is swept linearly in time~$\tau$. Hence, $\langle v_\ell \rangle\left[q(\tau) \right]$ is periodic in time with the Bloch period~$\tau_{\mathrm{Bloch}}$, Eq.~\eqref{eq:Bloch_period}. Its temporal average vanishes. This oscillation is called {\it Bloch oscillation} and Fig.~\ref{fig:blochoscillations}(a) shows its representation for the first Brillouin zone.

Bloch oscillations allow us to couple the momentum eigenstate~$\ket{p_0}$ of a free particle to the momentum eigenstate~$\ket{p_{n_p}=p_0+2 {n_p}\hbar k}$, by sequentially transferring $n_p$ times the momentum $2\hbar k$. The associated momentum transfer is the same as the one obtained when driving a sequence of $n_p^{\textrm{th}}$ first-order Bragg pulses as discussed in Sec.~\ref{sec:Multi-photon_Bragg}. For this purpose, $\ket{p_0}$ is adiabatically transferred to the lowest band of the Bloch lattice with $\ell=0$ by increasing the lattice depth $V_0$. In order to avoid inter-band transitions this transfer has to be adiabatic, that is the change $\diff V_0/\diff\tau$ of the potential amplitude~$V_0$, Eq. \eqref{eq:latticedepth}, has to be much smaller than $\Delta E^2/\hbar$, where $\Delta E $ denotes the energy difference between the bands depicted in Fig. \ref{fig:blochoscillations}(a), 

Atoms in the fundamental band are then coherently accelerated by applying a linear chirp~$2\pi\alpha$ to the frequency difference~$\delta\omega\equiv\omega_1-\omega_2$ of the two counterpropagating light waves with frequencies~$\omega_1$ and~$\omega_2$,
that is~\cite{Peik97PRA}
\begin{equation}\label{eq:frequency_chirp}
\delta\omega(\tau)\equiv 2\pi\alpha \tau\,.
\end{equation}
This chirp is equivalent to the effective force 
\begin{equation}\label{eq:bloch_force}
\left|F\right|=\frac{2\pi m}{k_\textrm{eff}} \left|\alpha\right|
\end{equation}
of Eq.~\eqref{eq:forceschroedinger} with~$k_{\textrm{eff}}\equiv 2 k$.

The accelerated lattice generated by this chirp is very well controllable and allows us to efficiently couple the momentum states~$\ket{p_0}$ and $\ket{p_{n_p}}$ by a sequential adiabatic transfer of the quasi-momentum~$\hbar q$ of the lattice to the atoms. Finally, the atoms are unloaded from the lattice by slowly decreasing the potential amplitude~$V_0$.

\subsubsection{Landau-Zener transitions}
In order to estimate the efficiency of driving Bloch oscillations and the associated momentum transfer, we have to take into account two main loss mechanisms: Spontaneous emission and inter-band transitions.

Although the detuning~$\Delta$ is large, still an appreciable fraction of atoms is lost due to spontaneous emission. We characterize the surviving fraction of atoms after the short acceleration time~$\tau_{\mathrm{acc}}$ by the parameter
\begin{equation}\label{eq:spontanfraction}
\eta_{\mathrm{sp}}\equiv 1-R_{\mathrm{sp}}\tau_{\mathrm{acc}}\,,
\end{equation}
where $R_{\mathrm{sp}}$ is defined by Eq.~\eqref{eq:spontan}.

\begin{figure}[h]
\includegraphics[width=\linewidth]{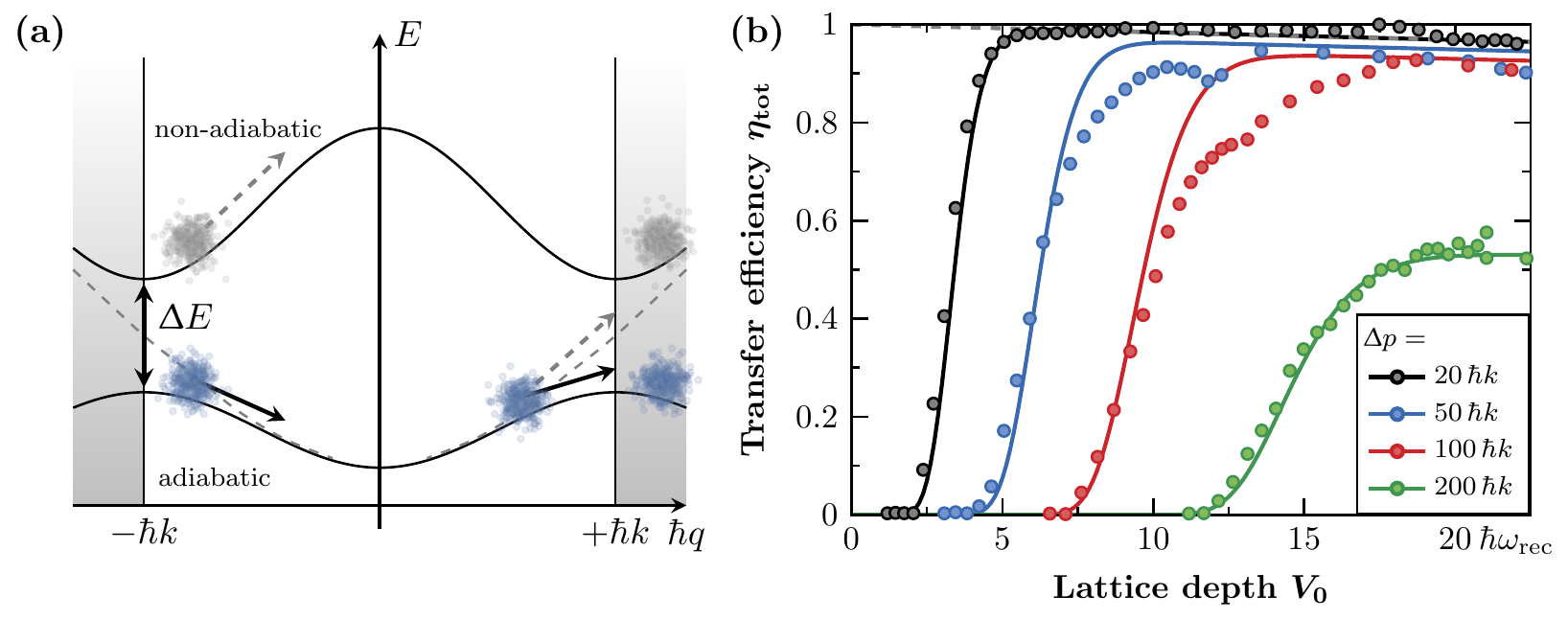}
\caption{Bloch oscillations in the first Brillouin zone (a) caused by a force $F$ acting on the center-of-mass motion of the atoms and transfer efficiency $\eta_{\mathrm{tot}}$ as a function (b) of the lattice depth~$V_0$ for the different momentum transfers~$\Delta p$. When the force is weak enough to avoid inter-band transitions, the atoms undergo an adiabatic acceleration for the time~$\tau_{\mathrm{acc}}=1$\,ms while they cross the individual Brillouin zones. Up to~$\Delta p=100\,\hbar k$ the fraction~$\eta_{\mathrm{tot}}$ of atoms remaining in the target momentum state 
can be held above 0.9. The solid lines are given by the product~$\eta_{\mathrm{tot}}=\eta_{\mathrm{sp}}\eta_{\mathrm{LZ}}$ normalized to the measured maximum transfer efficiency, where $\eta_{\mathrm{sp}}$ and $\eta_{\mathrm{LZ}}$ are defined by Eqs. \eqref{eq:spontanfraction} and \eqref{eq:landauzener}, accordingly. This figure is an adaptation of Figs.~4.11 and 5.23 in Ref.~\cite{Abend17}.}
\label{fig:blochoscillations}
\end{figure}

The second loss mechanism results from the fact that the lattice is not infinitely deep and the momentum transfer is not fast enough. These deficiencies give rise to inter-band transitions. Indeed, for a lattice generated by two counter-propagating light waves with the time-dependent frequency difference~$\delta\omega$ given by Eq.~\eqref{eq:frequency_chirp}, the chirp~$\alpha$ has to obey the adiabaticity criterion 
\begin{equation}\label{eq:adiabticcrit}
\left|\alpha\right|\equiv\left|\frac{\diff}{\diff \tau}\frac{\delta\omega}{2\pi}\right| =n_p\frac{\orec}{\tau_{\mathrm{acc}}}\ll \frac{\Delta E^2}{\hbar^2}.
\end{equation}
Here~$\Delta E$ denotes the energy of the band gap shown in Fig.~\ref{fig:blochoscillations}(a) and we aim at transferring $n_p$ pairs of photons during the time~$\tau_{\mathrm{acc}}$. 

We estimate the efficiency of the momentum transfer with the familiar Landau-Zener formula~\cite{Zener32}
\begin{equation}\label{eq:landauzener} 
\eta_{\mathrm{LZ}} = \left[1- \exp\left(-\frac{\pi}{2}\frac{\Delta E^2}{\hbar^2\alpha}\right)\right]^{n_p} \,,
\end{equation}
which provides us with the surviving fraction of atoms for a given chirp rate~$\alpha$ and band gap energy~$\Delta E$. The power~$n_p$ in Eq.~\eqref{eq:landauzener} indicating the number of transitions reflects the fact that an inter-band transition may take place at each crossing of the Brillouin zone, that is whenever a photon pair is scattered. 

The total fraction
\begin{equation}\label{eq:tot_fraction_surviving_atoms}
\eta_{\mathrm{tot}}\equiv \eta_{\mathrm{sp}}\eta_{\mathrm{LZ}}
\end{equation}
of surviving atoms after the acceleration is given by the product of $\eta_{\mathrm{sp}}$ and $\eta_{\mathrm{LZ}}$
 defined by Eqs.~\eqref{eq:spontanfraction} and \eqref{eq:landauzener}.
 
In Fig.~\ref{fig:blochoscillations}(b) we show by dots the measured efficiencies~$\eta_{\textrm{tot}}$ to transfer the momentum $\Delta p$ as a function of the lattice depth~$V_0$ for a fixed acceleration period~$\tau_{\mathrm{acc}}=1$\,ms. The solid lines are based on Eq.~\eqref{eq:tot_fraction_surviving_atoms} normalized to the measured maximum transfer efficiency. For~$\Delta p=20\,\hbar k$ and 200\,$\hbar k$ the solid line fits almost perfectly. However, for the data points with~$\Delta p=50\,\hbar k$ and 100\,$\hbar k$, residual resonant tunneling is visible, which is not taken into account in the Landau-Zener formula, Eq.~\eqref{eq:landauzener}. 

In principle, there are no fundamental limits on the amount~$\Delta p$ of momentum that can be transferred by Bloch oscillations during a fixed time. It is a purely technical issue. 

However, a limitation that is not easy to overcome is the spontaneous emission characterized by the rate~$R_{\mathrm{sp}}$, given by Eq.~\eqref{eq:spontan}. It effectively reduces the total fraction~$\eta_{\textrm{tot}}$ of atoms, since the laser detuning~$\Delta$ and the laser power~$P$ cannot be increased arbitrarily. For a laser detuning $\Delta=100$\,GHz, even at the largest lattice depth $V_0=23\, \hbar\omega_{\mathrm{rec}}$, the relative contribution of spontaneous emission still stays at the few percent level. Up to an acceleration of 100\,$\hbar k$/ms, the transfer efficiencies per photon $\eta_{\textrm{tot}}/\hbar k$ can be held at a level of $\eta_{\mathrm{tot}}/\hbar k>0.999$. However, for larger values of the acceleration, the relative transfer efficiency starts to decrease due to a violation of the adiabaticity criterion, Eq.~\eqref{eq:adiabticcrit}.

\subsection{Mach-Zehnder interferometer for gravity measurements}
\label{sec:Mach_Zehnder}

Measurements of gravity with atoms typically employ a MZI~\cite{Kasevich91PRL,Kasevich92APB,Peters99Nature,Peters01Metrologia,Borde01CRP,Storey,Schleich_NJP2013}     which is similar to its original optical analogue and consists of three elements: two beam splitters and one mirror. 

\subsubsection{Set-up}
\begin{figure}[h]
\includegraphics[width=\linewidth]{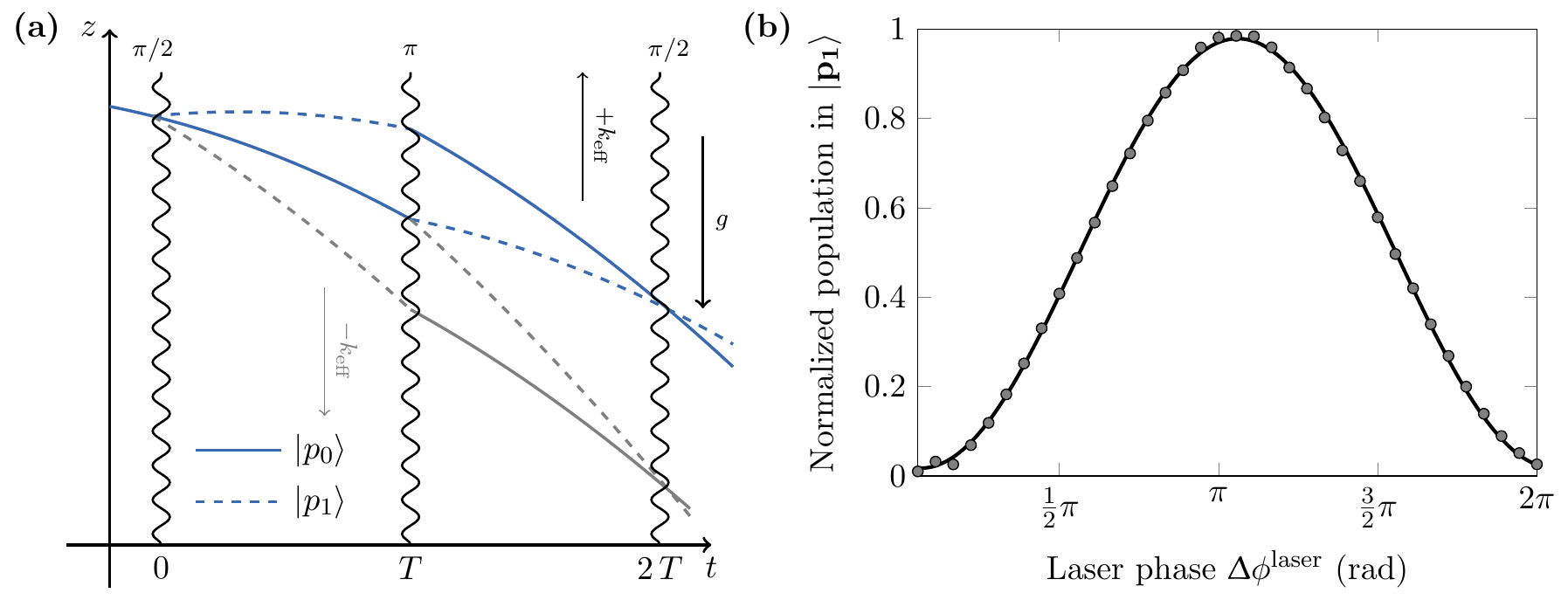}
\caption{Mach-Zehnder interferometer (MZI) for atoms. The spacetime diagram (a) shows the atomic trajectories due to a three-pulse-sequence consisting of (i) an initial $\pi/2$-pulse to prepare a superposition of the two different momentum states $\ket{p_0}$ and $\ket{p_1}$ of the atomic center-of-mass motion, (ii)  a $\pi$-pulse to exchange the imprinted momenta, and (iii) a final $\pi/2$-pulse to obtain the interference in the atomic population in an exit port. In (b) we depict a fringe scan for an adjusted laser phase~$\Delta\phi^{\mathrm{laser}}$ in the interval~$[0,2\pi]$. Figure (a) is reproduced from Fig.~4.15(b) in Ref.~\cite{Abend17}.}
\label{fig:machzehnder}
\end{figure}

The first beam splitter generates a coherent superposition of two different momentum states~$\ket{\vec{p}_0}$ and $\ket{\vec{p}_1}$ of the atomic center-of-mass motion. This superposition results in two spatially separated trajectories associated with each of the two momentum states, as depicted in Fig.~\ref{fig:machzehnder}(a). After a time~$T$, a mirror exchanges these two momentum states, leading to a redirection of the trajectories. Finally, after a total duration of~$2T$, the two trajectories overlap again, and the second beam splitter is applied. 

As discussed in Sec.~\ref{sec:Rabi_oscillations} we describe beam splitters and mirrors realized by the atom-light interaction in the Rabi formalism, and the corresponding pulse sequence for the MZI reads $\pi/2-\pi-\pi/2$. Here the two beam splitters ($\pi/2$-pulse) are separated from the mirror ($\pi$-pulse) by~$T$. Usually, the momentum transfer $\hbar \keffvc$ induced by a beam splitter or mirror is parallel or antiparallel to the gravitational acceleration $\boldsymbol{g}$ as shown in Fig.~\ref{fig:machzehnder}(a).

The interference signal of the interferometer is determined by the atomic population 
\begin{equation}\label{eq: interferometer_output_probability}
P (\Delta \phi)=\frac{1}{2}\left[1-C \cos(\Delta \phi)\right]
\end{equation} 
in the momentum state~$\ket{\vec{p}_1}$ after the final beam splitter, where $C$ and $\Delta \phi$ are the contrast and the total phase shift, respectively.

In the remainder of these lectures we consider only a perfectly closed interferometer, where the two wave packets propagated along the two trajectories perfectly overlap after the total interferometer time~$2T$. In this case we obtain the maximal contrast $C=1$.

\subsubsection{Contributions to phase shift}
The total phase shift~$\Delta \phi$ of the MZI reads~\cite{KLEINERT20151,Louchet11NJP}
\begin{equation}\label{eq:ouputphase}
\Delta \phi=\Delta \phi^{\mathrm{laser}}+\Delta \phi^{\mathrm{ac}}+\Delta \phi^{\mathrm{2ph}} + \Delta \phi^{\mathrm{inert}}\,,
\end{equation}
where the laser phase
\begin{equation}\label{eq:laserphase} 
\Delta \phi^{\mathrm{laser}}\equiv\phi_1-2\phi_2+\phi_3
\end{equation}
is determined by the laser phases~$\phi_1$, $\phi_2$, and $\phi_3$ imprinted on the atom wave during the atom-light interaction due to the first, second, and third pulse. The phase~$\Delta\phi^{\mathrm{laser}}$ is of the form of a discrete second derivative. Unless~$\phi_3$ is used to modulate the signal, as depicted in Fig.~\ref{fig:machzehnder}(b), the laser phase~$\Delta\phi^{\mathrm{laser}}$ vanishes.

The single- and two-photon light shifts~$\Delta \phi^{\mathrm{ac}}$ and~$\Delta \phi^{\mathrm{2ph}}$ may lead to an offset shift, which in the first order depends on the difference in phase imprinted during the first and last pulse. In contrast to Raman diffraction, where the ratio of the intensities of the two frequency components needs to be properly adjusted~\cite{Berg14,Schlippert14}, the phase $\Delta\phi^{\mathrm{ac}}$ can be intrinsically suppressed in interferometers based on Bragg diffraction~\cite{Giese16PRA}. 

The two leading-order contributions to the phase shift
\begin{equation}
\Delta \phi^{\mathrm{inert}}\equiv\Delta \phi^{\mathrm{grav}}+\Delta \phi^{\mathrm{rot}}
\end{equation}
containing inertial effects, originate from the gravitational acceleration~$\boldsymbol{g}$ 
\begin{equation}\label{eq:phasegravity}
\Delta \phi^{\mathrm{grav}} \equiv\vec{\keff}\cdot\boldsymbol{g} \,T^2,
\end{equation}
and from the rotation of the Earth with frequency~$\vec{\Omega}_{\mathrm{E}}$
\begin{equation}\label{eq:phasesagnac}
\Delta \phi^{\mathrm{rot}} \equiv 2 \vec{\keff}\cdot(\vec{\Omega}_{\mathrm{E}} \times \vec{v}_0)T^2,
\end{equation}
representing the Sagnac effect.
Here~$\vec{v}_0$ is the velocity of the atoms before the first beam splitter. 

\subsubsection{Influence of non-zero pulse duration}
If the duration~$\tau_{\mathrm{p}}$ of a pulse is not negligibly short compared to the pulse separation time~$T$, a modification of~$\Delta\phi^{\mathrm{grav}}$ given by Eq.~\eqref{eq:phasegravity} is necessary~\cite{Antoine07PRA,Cheinet08IEEE,Stoner11JOSAB,Bonnin15PRA}. For a Gaussian-shaped pulse of width $\sigma_{\tau_\mathrm{p}}$ we find the new pulse separation time
\begin{equation}
T'\equiv T+\tau_\mathrm{p}-\tau'_\mathrm{p}\,,
\end{equation}
where 
\begin{equation}
\tau'_{\mathrm{p}}\equiv \sqrt{2\pi}\sigma_{\tau_\mathrm{p}}
\end{equation}
is the duration of an equivalent box-shaped pulse covering the same area as the Gaussian-shaped one. Here we have assumed equally long $\pi$- and $\pi/2$-pulses, which, however, differ in intensity. 

As a result the improved expression 
\begin{equation}
\Delta\phi^{\mathrm{grav}} = \keffvc \cdot \boldsymbol{g}\,  T'^2\left[1+\left(1+\frac{2}{\pi}\right)\frac{\tau'_{\mathrm{p}}}{T'}+...\right]\,,
\end{equation}
for~$\Delta\phi^{\mathrm{grav}}$ represents a Taylor expansion in powers of $\tau_{\mathrm{p}}'/T'$.

Hence, the influence of the correction due to a non-zero pulse duration decreases for larger~$T$, since the duration~$\tau_{\mathrm{p}}$ is independent of the separation~$T$.

\subsubsection{Measurement of gravitational acceleration}
\label{sec:Chirping_laser_frequency}
The phase~$\Delta\phi$ accumulated in an atom interferometer does not only depend on the laser phase~$\Delta\phi^\mathrm{laser}$, but also on the time dependence of the laser frequency~$\Delta\nu=\Delta \nu(\tau)$. In the following we restrict ourselves to a one-dimensional problem with the gravitational acceleration $g=\left|\boldsymbol{g}\right|$, and neglect for the time being the phase contributions~$\Delta\phi^\mathrm{laser}$ and~$\Delta\phi^\mathrm{rot}$. 

During the free fall of the atom the resonance frequency of the transition changes due to the Doppler effect, Eq.~\eqref{eq:dopplerfrequency}. This effect needs to be taken into account by a proper change of the detuning~$\Delta \nu(\tau)$ after a certain free-fall time~$\tau$. 

Indeed, the appropriate time dependence 
\begin{equation}
\Delta \nu(\tau)=\frac{\delta\omega(\tau)}{2\pi}
\end{equation}  
of the chirp originates from the requirement for~$\delta \omega$ to stay on resonance, Eq.~\eqref{eq:seqtrans}, where the Doppler frequency~$\omega_\mathrm{D}=\omega_\mathrm{D}(0)$, Eq.~\eqref{eq:dopplerfrequency}, is replaced by
\begin{equation}
\omega_\mathrm{D}(\tau)=\omega_\mathrm{D}(0)-\keff g \tau\,.
\end{equation}

Chirping the laser frequency is not only a necessity to remain on resonance, but the chirp rate~$\alpha$ also defines the acceleration $a=2\pi \alpha/\keff$, compare to Eq.~\eqref{eq:bloch_force}, of the wavefronts of the Bragg lattice during free fall. In particular, the adjustment of~$2\pi \alpha$ can be used to modulate the interferometer phase~$\Delta\phi$ in a way that
\begin{equation} \label{eq:chripratemeasurement} 
\Delta \phi(\alpha,T) = \left(g-\frac{2\pi\alpha}{\keff}\right)\keff T^2 \mathrm{.}
\end{equation}

It is the control of this chirp rate that allows us to realize an atomic gravimeter. Indeed, Eq.~\eqref{eq:chripratemeasurement} relates the output of the atom interferometer~$P(\Delta\phi)$, Eq. \eqref{eq: interferometer_output_probability}, depending on the gravitational acceleration~$g$, to the two well-controlled parameters~$T$ and~$\alpha$. To extract~$g$, we evaluate the dependence of the output signal~$P(\Delta\phi)$  on one parameter with the other one fixed. 

\begin{figure}[h]
\includegraphics[width=\linewidth]{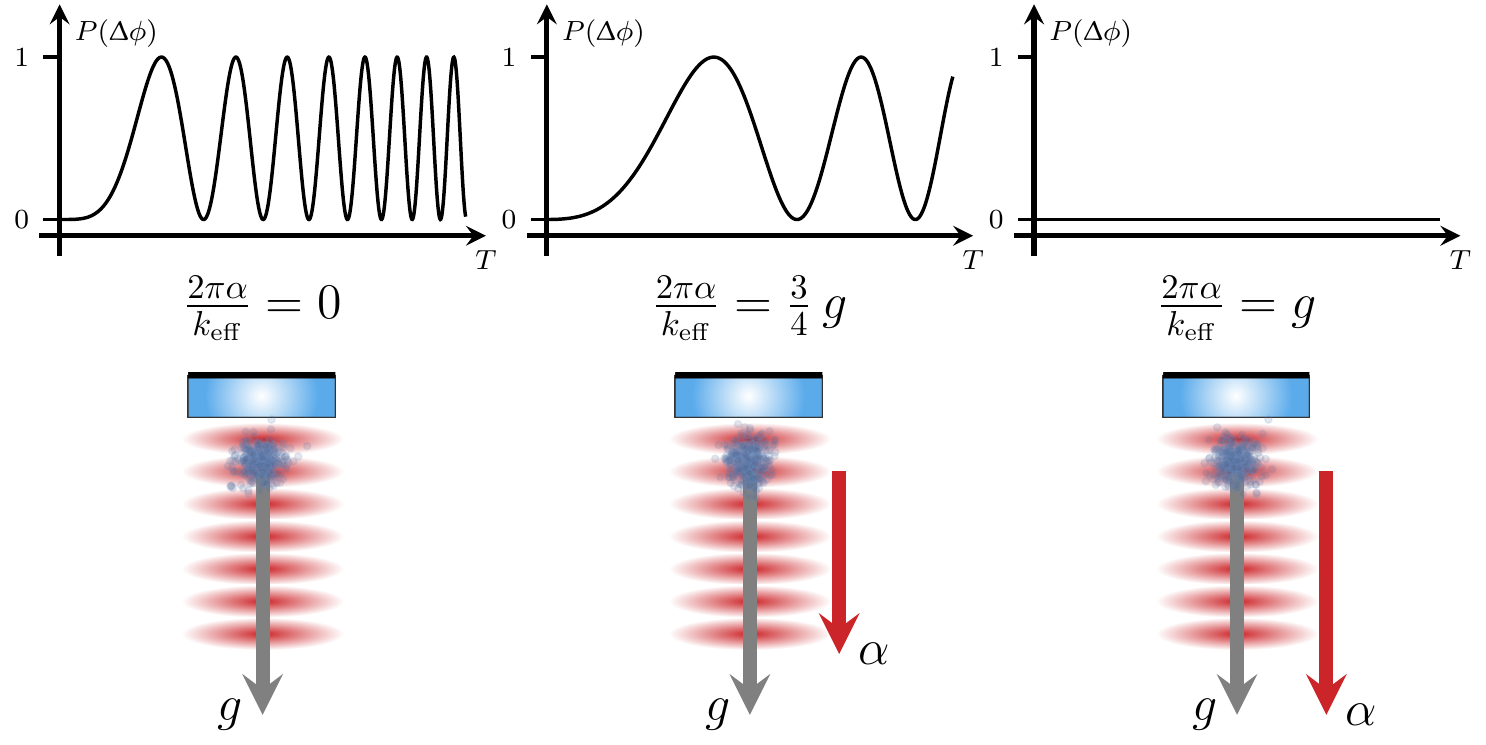}
\caption{Local gravitational acceleration~$g$ obtained from the measurement of the chirp rate in an atomic gravimeter. 
The number of atoms in the exit port of an atom interferometer governed by~$P=P(\Delta\phi)$ is counted for different chirp rates~$\alpha$ of the laser frequency displayed here for the three lattice wavefront accelerations $2\pi\alpha/\keff=0,\;3g/4,$ and $g$. 
For increasing pulse separation time~$T$, a chirped sinusoidal oscillation is obtained. The chirp rate decreases
with increasing~$\alpha$ and finally vanishes for $2\pi\alpha/\keff=g$. 
This picture is an adaptation of Fig.~2.4 in Ref.~\cite{Schlippert14} and Fig.~4.19 in Ref.~\cite{Abend17}.}
\label{fig:chirpratemeasurement}
\end{figure}

Figure~\ref{fig:chirpratemeasurement} displays~$P=P(\Delta\phi)$ versus the variation of the pulse separation time~$T$ for three different values of~$2\pi\alpha/\keff$, namely $0,\,3g/4,$ and $g$. The closer the ratio~$2\pi\alpha/\keff$ is to the gravitational acceleration~$g$, the slower is the oscillation chirp, and for~$2\pi\alpha/\keff=g$ the oscillations vanish completely. 

When there is a significant missmatch between the chirp rate~$\alpha$ and the free-fall rate~$g$, a reduction in excitation efficiency due to a non-compensated Doppler shift appears. This effect was neglected in Eq.~\eqref{eq:chripratemeasurement} and Fig.~\ref{fig:chirpratemeasurement}. 

We conclude by noting that in a real experiment the ratio~$2\pi\alpha/\keff$ differs from the gravitational acceleration by less than one percent.

\section{Equivalence principle and atom interferometry}
\label{sec:EQP}

A prime application of an inertially sensitive atom interferometer is its use as a probe of Einstein's equivalence principle (EEP). We devote the present section to this topic and emphasize that the results summarized therein have originally been published in Ref.~\cite{Schlippert14PRL}. 

First, we outline in Sec.~\ref{sec:UFF_test} a framework for testing the universality of free fall (UFF). We then present in Sec.~\ref{sec:Simultaneous_interferometer} an experiment based on a dual-species rubidium and potassium interferometer using atoms released from an optical molasses together with Raman diffraction to probe the UFF. Finally, we summarize our results and perform the associated data analysis in Sec.~\ref{sec:UFF_data_analysis}.

\subsection{Frameworks for tests of the universality of free fall}
\label{sec:UFF_test}

In spite of being 36 orders of magnitude weaker than the Coulomb interaction, gravitation dominates on a cosmological scale. Since astrophysical objects are electrically neutral gravitation governs the structure of our universe.
Einstein's metric theory of gravity~\cite{Einstein16AP}, that is general relativity, provides us with the tools necessary to understand a vast variety of astronomical phenomena and makes verifiable predictions, such as the existence of gravitational waves \cite{Ref:Gravity1,Ref:Gravity2}. As of today, however, a completely satisfactory microscopic theory of {\it quantum} gravity merging general relativity with quantum mechanics is still lacking.

The EEP is a cornerstone of general relativity, and unification attempts in general imply the violation of at least one of its three central assumptions: i) Local position invariance, ii) local Lorentz invariance, and iii) the UFF.
Ranging from the famous Pound-Rebka experiment~\cite{Pound60PRL} and gravity probe A~\cite{Vessot80PRL}, to the extremely sensitive torsion balance experiments~\cite{Schlamminger08PRL} and lunar laser ranging~\cite{Williams04PRL,Williams12CQG,Mueller12CQG}, the EEP has been tested extensively.

Experiments employing matter wave interferometry have recently extended the landscape of classical UFF tests by entering the quantum domain.
The UFF postulates the equality of inertial mass $m_{\text{in}}$ and gravitational mass $m_{\text{gr}}$.
Given two bodies A and B it can be tested by obtaining the E\"otv\"os ratio  
\begin{equation}\label{eq:eotvos}
\eta_{\,\text{A,B}}\equiv 2\thickspace\frac{g_{\text{A}}-g_{\text{B}}}{g_{\text{A}}+g_{\text{B}}}
=2\thickspace\frac{\left(\frac{m_{\text{gr}}}{m_{\text{in}}}\right)_{\text{A}}
	-\left(\frac{m_{\text{gr}}}{m_{\text{in}}}\right)_{\text{B}}}
{\left(\frac{m_{\text{gr}}}{m_{\text{in}}}\right)_{\text{A}}
	+\left(\frac{m_{\text{gr}}}{m_{\text{in}}}\right)_{\text{B}}}
\end{equation}
for their respective gravitational accelerations $g_{\text{A}}$ and $g_{\text{B}}$.

Any non-zero measurement of $\eta_{\,\text{A,B}}$ would imply a composition-dependent inequality of inertial and gravitational mass. Indeed, the values
\begin{equation}
\eta_{\,\text{Earth,Moon}}=(-0.8\pm 1.3)\cdot 10^{-13}
\end{equation}
based on Lunar-Laser-Ranging~\cite{Williams04PRL,Williams12CQG,Mueller12CQG} and
\begin{equation}
\eta_{\,\text{Be,Ti}}=(0.3\pm 1.8)\cdot 10^{-13}
\end{equation}
employing torsion balances~\cite{Schlamminger08PRL} have provided the best constraints on violations of the UFF for a number of years.

Alternatively, one can also compare pairs of freely-falling test masses. The best result on ground~\cite{Niebauer87PRL} corresponds to 
\begin{equation}
\eta_{\,\text{Cu,U}}=(1.3\pm 5.0)\cdot 10^{-10}.
\end{equation}
Moreover, a recent test in space~\cite{Touboul12CQG,Touboul2017PRLMICROSCOPE} obtained
\begin{equation}
\eta_{\,\text{Ti,Pt}}=\left[-0.1\pm 0.9(\mathrm{stat})\pm0.9(\mathrm{syst})\right]\cdot 10^{-14}
\end{equation}
corresponding to an order of magnitude improvement of the bounds set by previous experiments.

The tests listed above employ classical test masses.
In analogy to the first observation of a gravitation-induced phase in a neutron interferometer~\cite{Colella75PRL}, the acceleration measured by the interference of a quantum object can be compared to that of a classical object~\cite{Peters99Nature,Merlet10Metrologia}, or the differential gravitational phase shifts between two quantum objects~\cite{Fray04PRL,Fray09SSRev,Bonnin13PRA,Tarallo14PRL,Zhou15PRL} can be exploited to search for violations of the UFF.
Beyond the comparison of two isotopes of strontium in 2014~\cite{Tarallo14PRL}, the latter approach was extended to comparing the free-fall acceleration of the two different elements rubidium and potassium~\cite{Schlippert14PRL}.

Experiments using quantum objects are beneficial because they are generally subject to systematic effects different from those dominating classical tests.
Moreover, unique properties of quantum objects, such as a macroscopic coherence length~\cite{Goeklue08CQG} and spin polarization~\cite{Leitner64PRL,Laemmerzahl98,Laemmerzahl98proceedings,Laemmerzahl06APB} can be tested as features possibly coupling to EEP-violating effects.
Furthermore, the set of available test masses is enhanced by those that can be laser-cooled and chosen as to maximize the sensitivity to violations.
This approach can be clearly illustrated by dilaton models~\cite{Damour12CQG}, and the so-called Standard Model Extension (SME)~\cite{Kostelecky11,Hohensee13PRL}, which both provide consistent frameworks for parametrizing violations of the EEP.

Violations of the UFF can be naturally parameterized by writing the acceleration~$g_{\text{X}}$ of a species X as 
\begin{equation}
g_{\text{X}}=\left(1+\beta_{\text{X}}\right)g\,,
\end{equation}
where $\beta_{\text{X}}$ is a small parameter which is species dependent, and vanishes in the absence of UFF violations. 

For dilaton models one has
\begin{equation}
\beta_{\text{X}}=D_1 Q^{'1}_{\text{X}} +D_2 Q^{'2}_{\text{X}}\,,
\end{equation}
where $D_1$ and $D_2$ are fundamental violation parameters, whereas \Q{X} and \QQ{X} are effective charges that mainly depend on the proton and neutron numbers of species X~\cite{Damour12CQG}.

The E\"otv\"os ratio for two species A and B is then given by
\begin{equation}
\eta_{\,\text{A,B}}\approx \beta_{\text{A}}-\beta_{\text{B}}=D_1(Q^{'1}_{\text{A}}-Q^{'1}_{\text{B}})+D_2(Q^{'2}_{\text{A}}-Q^{'2}_{\text{B}})\,,
\end{equation}
and the differences of the effective charges, which determine the sensitivity to violations associated with $D_1$ or $D_2$ are listed in Table \ref{tab:Damour} for several test pairs. 

Similarly, for the SME one has 
\begin{equation}
\beta_{X}= \thickspace f_{\beta^{e+p-n}_{\text{X}}}\beta^{e+p-n}+f_{\beta^{e+p+n}_{\text{X}}}\beta^{e+p+n}
+f_{\beta^{\bar{e}+\bar{p}-\bar{n}}_{\text{X}}}\beta^{\bar{e}+\bar{p}-\bar{n}}
+f_{\beta^{\bar{e}+\bar{p}+\bar{n}}_{\text{X}}}\beta^{\bar{e}+\bar{p}+\bar{n}}\,,
\end{equation}
where $\beta^{e+p-n}$, $\beta^{e+p+n}$, $\beta^{\bar{e}+\bar{p}-\bar{n}}$, and $\beta^{\bar{e}+\bar{p}+\bar{n}}$ parametrize the violations for various weighted combinations of elementary particles. Moreover, the sensitivity factors \fbminus{X} (\fbbarminus{X}) and \fbplus{X} (\fbbarplus{X}) are charges related to the neutron excess, and the overall baryon number in a given normal matter (antimatter) nucleus~\cite{Hohensee13PRL}.

\begin{table}[t!]
\centering
\small\renewcommand{\arraystretch}{1.4}

\begin{center}
\begin{tabular}{|l|c|c|c|c|}
\hline
A& B&Ref.&(\Q{A}\spaceminus\Q{B})$\cdot 10^4$&(\QQ{A}\spaceminus\QQ{B})$\cdot 10^4$\\
\hline
\textsuperscript{9}Be& Ti&\cite{Schlamminger08PRL}  &$-15.46$ &$-71.20$\\

Cu& \textsuperscript{238}U&\cite{Niebauer87PRL}  & $-19.09$ &$-28.62$\\

\Rbb&\Rb&\cite{Fray04PRL,Bonnin13PRA} &$0.84$& $-0.79$\\

\textsuperscript{87}Sr&\textsuperscript{88}Sr &\cite{Tarallo14PRL} &$0.42$& $-0.39$ \\

\textsuperscript{6}Li&\textsuperscript{7}Li~\footnotemark[1] &\cite{Hamilton12DAMOP}&$0.79$& $-10.07$\\
\hline
\K&\Rb&\cite{Schlippert14PRL}& $-6.69$& $-23.69$\\

\hline
\end{tabular}
\end{center}

\caption[Comparison model of test masses \textup{A} and \textup{B} in the dilaton]{Comparison of test masses \textup{A} and \textup{B} analyzed in the dilaton model. The charges 
\Q{\textup{X}}~and \QQ{\textup{X}}~with \textup{X} being either \textup{A} or \textup{B} are calculated according to Ref.~\cite{Damour12CQG}. 
A larger absolute number corresponds to a larger anomalous acceleration, and thus a higher sensitivity to violations of the \textup{EEP}. 
For \textup{Ti} and \textup{Cu} natural occurrence of isotopes is assumed. This table is a reproduction of Table~2.1 in Ref.~\cite{Schlippert14}.}
\label{tab:Damour}
\end{table}
\footnotetext[1]{A UFF test comparing \textsuperscript{6}Li {\it vs.} \textsuperscript{7}Li has not yet been performed.}

The corresponding E\"otv\"os ratio reads
\begin{equation}\label{eq:Damour}
\eta_{\,\text{A,B}}\approx\beta_{\text{A}}-\beta_{\text{B}}= ~\Delta f_{-n}\beta^{e+p-n}+\Delta f_{+n}\beta^{e+p+n}+\bar{\Delta f_{-n}}\beta^{\bar{e}+\bar{p}-\bar{n}}+\bar{\Delta f_{+n}}\beta^{\bar{e}+\bar{p}+\bar{n}}
\end{equation}
with
\begin{equation}
\begin{aligned}
\Delta f_{-n} \equiv f_{\beta^{e+p-n}_{\text{A}}}- f_{\beta^{e+p-n}_{\text{B}}}\\
\Delta f_{+n} \equiv f_{\beta^{e+p+n}_{\text{A}}}- f_{\beta^{e+p+n}_{\text{B}}}\\
\bar{\Delta f_{-n}} \equiv f_{\beta^{\bar{e}+\bar{p}-\bar{n}}_{\text{A}}} - f_{\beta^{\bar{e}+\bar{p}-\bar{n}}_{\text{B}}}\\
\bar{\Delta f_{+n}} \equiv f_{\beta^{\bar{e}+\bar{p}+\bar{n}}_{\text{A}}} -f_{\beta^{\bar{e}+\bar{p}+\bar{n}}_{\text{B}}}\text{.}
\end{aligned}
\end{equation}
These differences of sensitivity factors are listed in Table \ref{tab:SME} for relevant test pairs.

Tables \ref{tab:Damour} and \ref{tab:SME} clearly show that the choice of test masses heavily influences the achievable impact on global violation bounds.
Specifically, this information allows us to use novel test pairs that were previously unconstrained, or only weakly constrained, to improve the bounds on certain violation parameters~\cite{Hohensee13PRL,Mueller13varenna}.
Here, a new, and independent result can have an enormous impact on the global model even if it does not reach the state-of-the-art sensitivity.

\begin{table}[ht]
\centering
\small\renewcommand{\arraystretch}{1.4}

\begin{center}
\begin{tabular}{|l|c|c|c|c|c|c|}
\hline
A& B&Ref.&$\Delta f_{-n}\cdot 10^2$&$\Delta f_{+n}\cdot 10^4$&$\bar{\Delta f_{-n}}\cdot 10^5$&$\bar{\Delta f_{+n}}\cdot 10^4$\\
\hline
\textsuperscript{9}Be& Ti&\cite{Schlamminger08PRL}  & $1.48$ &$-4.16$ & $-0.24$ &$-16.24$\\

Cu& \textsuperscript{238}U&\cite{Niebauer87PRL}  & $-7.08$ &$-8.31$ & $-89.89$ &$-2.38$\\

\Rbb&\Rb&\cite{Fray04PRL,Bonnin13PRA} &$-1.01$& $1.81$ &$1.04$& $1.67$\\

\textsuperscript{87}Sr&\textsuperscript{88}Sr &\cite{Tarallo14PRL}&$-0.49$& $2.04$ &$0.81$& $1.85$\\
\textsuperscript{6}Li&\textsuperscript{7}Li~\footnotemark[2] &\cite{Hamilton12DAMOP}&$-7.26$& $7.79$ &$-72.05$& $5.82$\\
\hline
\K&\Rb&\cite{Schlippert14PRL}& $-6.31$& $1.90$& $$-62.30$$&$ 0.64$\\
\hline

\end{tabular}
\end{center}

\caption[Comparison of test masses \textup{A} and \textup{B} in the SME]{Comparison of test masses \textup{A} and \textup{B} analyzed in the Standard Model Extension. 
The sensitivity factors $\Delta f_{-n}$, $\Delta f_{+n}$, $\bar{\Delta f_{-n}}$, and $\bar{\Delta f_{+n}}$~are calculated according to Ref.~\cite{Hohensee13PRL}. 
Relevant nuclide data is taken from Ref.~\cite{Audi03}. 
A larger absolute number corresponds to a larger anomalous acceleration, and thus higher sensitivity to violations of the \textup{EEP}. 
For \textup{Ti} and \textup{Cu} natural occurrence of isotopes~\cite{Laeter09} is assumed. This table is a reproduction of Table~2.2 in Ref.~\cite{Schlippert14}.}
\label{tab:SME}
\end{table}
\footnotetext[2]{A UFF test comparing \textsuperscript{6}Li {\it vs.} \textsuperscript{7}Li has not yet been performed.}

\subsection{Simultaneous \textup{\Rb}~and \textup{\K}~interferometer}
\label{sec:Simultaneous_interferometer}

\begin{figure}[ht]
	\centering
\includegraphics[width=0.5\textwidth]{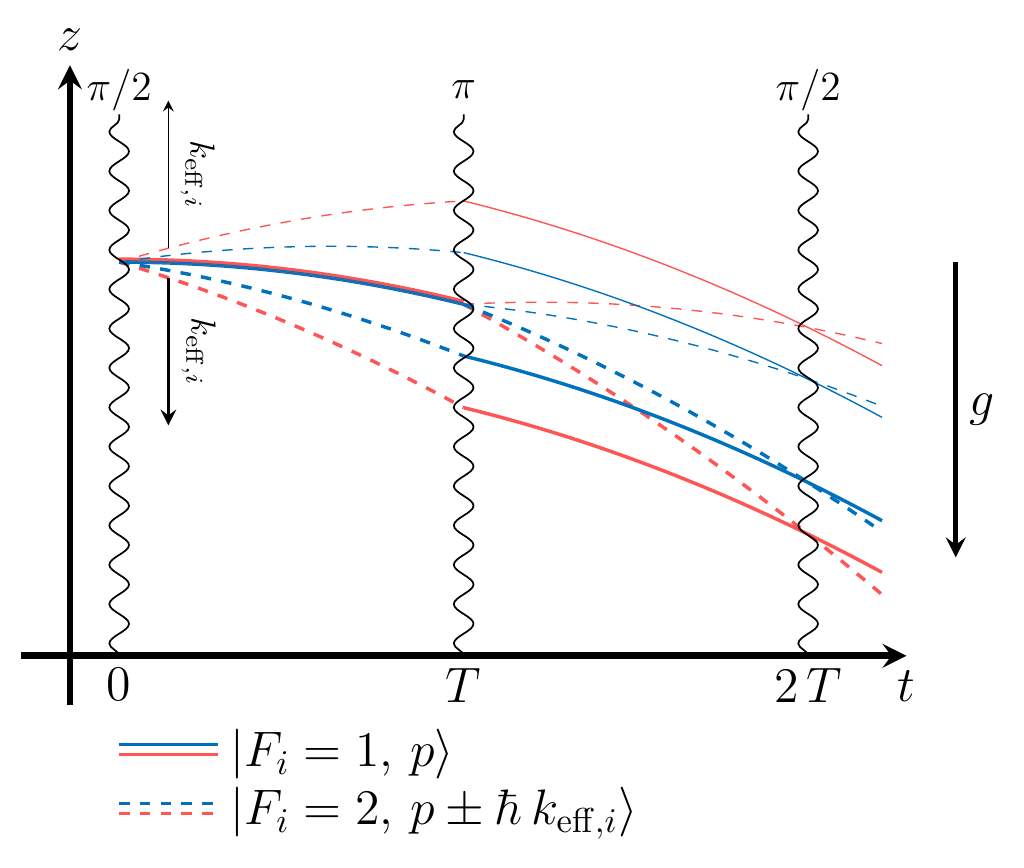}
\caption[]{Spacetime diagram of a dual-species {Mach-Zehnder} atom interferometer in a constant gravitational field for the downward (thick lines) and upward (thin lines) direction of the momentum transfer. Stimulated {Raman} transitions at times $0$, $T$, and $2\,T$ couple the states $\ket{F_i=1,\,p}$ and $\ket{F_i=2,\,p\pm\hbar\,\keffi{i}}$, where $i$ stands for Rb (blue lines) or K (red lines). The velocity change induced by the {Raman} pulses is not to scale compared to the gravitational acceleration. This figure is reproduced from Fig.~5.1 in Ref.~\cite{Schlippert14}.}
\label{fig:mzdual}
\end{figure}

\begin{figure}[h!]
	\centering
	\includegraphics[width=0.95\textwidth]{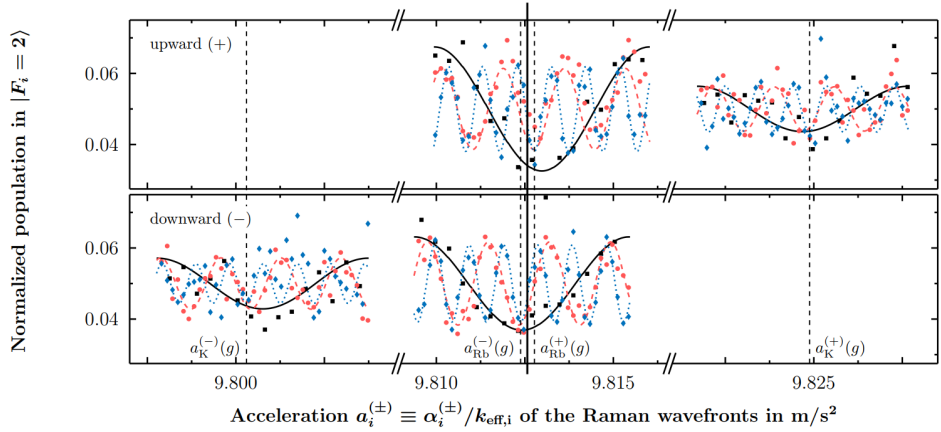}
	\caption[Determination of the differential gravitational acceleration of rubidium and potassium]{Determination of the differential gravitational acceleration of rubidium and potassium. Typical interference signals and sinusoidal fits as a function of the effective Raman wavefront acceleration are shown for pulse separation times $T=\SI{8}{ms}$ (black squares and solid black line), $T=\SI{15}{ms}$ (red circles and dashed red line), and $T=\SI{20}{ms}$ (blue diamonds and dotted blue line) both for the upward $(+)$ and downward $(-)$ directions of the momentum transfer. The central fringe positions $a_i^{(\pm)}(g)$ (dashed vertical lines), where $i$ is Rb or K, are shifted symmetrically around $g_i=[a_i^{(+)}(g)-a_i^{(-)}(g)]/2$ (solid vertical line). The data sets are corrected for slow linear drifts and offsets. This figure is reproduced from Fig.~5.2 in Ref.~\cite{Schlippert14}.}
	\label{fig:qtufffringes}
\end{figure}

To test the UFF, we operate two {Mach-Zehnder}-type matter wave interferometers~\cite{Kasevich91PRL} with laser-cooled \Rb~and \K~as shown in Fig.~\ref{fig:mzdual}.
To leading order, the phase shift in each interferometer reads
\begin{equation}\label{eq:phaseshift}
\Delta\phi_i=\left(g_i-\frac{\alpha_i}{\keffi{i}}\right)\keffi{i} T^2\,,
\end{equation}
and can be deduced by counting the atoms in the exit ports. Here the effective wavefront acceleration  $\alpha_i/\keffi{i}$  introduced by linearly ramping the Raman laser frequency difference is utilized to null the phase shift induced by the gravitational acceleration. We emphasize that the definition of the chirp rate $ \alpha_i$ differs by a factor of $2\pi$ from the definition of the chirp rate $\alpha$ in Sec.~\ref{sec:Chirping_laser_frequency}.

An ensemble of $8\cdot 10^8$ atoms ($3\cdot 10^7$ atoms) of rubidium (potassium) atoms is collected from a two-dimensional magnetooptical trap, then sub-Doppler cooled~\cite{Landini11PRA,Chu98,Phillips98} to $T_{\text{Rb}}=\SI{27}{\micro K}$ ($T_{\text{K}}=\SI{32}{\micro K}$), and after being optically pumped to the $\ket{F_i=1}$ manifold, released into free fall.
Three two-photon Raman pulses separated in time by $T$ are applied to coherently split, redirect, and recombine the atoms during free fall.
For detection, the population in $\ket{F_i=2}$ as well as the total population in $\ket{F_i=1}$ and $\ket{F_i=2}$ are obtained via state-selective fluorescence detection, yielding the normalized excitation probability.
The overall cycle time is about~$\SI{1.6}{s}$.

A global minimum of the interference fringes appears independently of the free evolution time $T$, when the condition $g_i-\alpha_i/\keffi{i}=0$ is fulfilled. We display this concept for the upward and downward directions of the momentum transfer\footnote[3]{The differential signal of the upward and downward directions of the momentum transfer allows us to suppress~\cite{McGuirk02PRA,Louchet11NJP} spurious phase shifts independent of the direction of $\keff$.} in Fig.~\ref{fig:qtufffringes}.

\subsection{Data analysis and result}
\label{sec:UFF_data_analysis}
Over a duration of about four hours we have tracked the central fringe position $a_i^{(\pm)}(g)$ by scanning across the minimum in 10 steps with alternating directions of the momentum transfer at a pulse separation time of $T=\SI{20}{ms}$.
Two scans yield $g_i=[a_i^{(+)}(g)-a_i^{(-)}(g)]/2$ and can be used to compute the E\"otv\"os ratio, Eq.~\eqref{eq:eotvos}.

Systematic effects affecting our measurement are listed in Table~\ref{tab:systematics}.
The total bias $\Delta\eta_{\text{tot}}=-5.4\cdot 10^{-8}$ is subject to an uncertainty $\delta\eta_{\text{tot}}=3.1\cdot 10^{-8}$. Considering the systematic and statistical uncertainty as well as the bias from Table~\ref{tab:systematics}, an overall result of 
\begin{equation}
\eta_{\mathrm{Rb},\mathrm{K}}=\etav
\end{equation}
is obtained.

The column $\delta\eta^{\text{adv}}$ points to possible future improvements (indicated in bold face) using a common optical dipole trap~\cite{Zaiser11PRA} as a source to further cool and collocate rubidium as well as potassium, and gain better control over their initial conditions.

\begin{table}[h]
\centering
\small\renewcommand{\arraystretch}{1.4}

\begin{center}
\begin{tabular}{|l|c|c|c|}
\hline
Contribution & \EotDelta &\Eotsigma & $\delta\eta^{\text{adv}}$ \\
\hline

Second-order Zeeman effect& $-5.8\cdot 10^{-8}$ &$2.6\cdot 10^{-8}$&$\mathbf{3.0\cdot 10^{-9}}$\\

Wavefront aberration & 0 &$1.2\cdot 10^{-8}$&$\mathbf{3.0\cdot 10^{-9}}$\\

Coriolis force& 0 &$9.1\cdot 10^{-9}$&$\mathbf{1.0\cdot 10^{-11}}$\\

Two-photon light shift& $4.1\cdot 10^{-9}$ &$8.2\cdot 10^{-11}$&$8.2\cdot 10^{-11}$\\

Effective wave vector& 0 &$1.3\cdot 10^{-9}$&$1.3\cdot 10^{-9}$\\

First-order gravity gradient& 0 &$9.5\cdot 10^{-11}$&$\mathbf{1.0\cdot 10^{-12}}$\\
\hline
Total &$-5.4\cdot 10^{-8}$&$3.1\cdot 10^{-8}$&$\mathbf{4.4\cdot 10^{-9}}$\\
\hline

\end{tabular}
\end{center}
\caption{Systematic biases \EotDelta~and comparison between the uncertainties $\delta\eta$ and $\delta\eta^{\textup{adv}}$ of the E\"otv\"os ratio in the current, and in an advanced set-up. The improved values highlighted in bold face arise from the use of an optical dipole trap. The uncertainties are assumed to be uncorrelated at the level of the inaccuracy. This table is a reproduction of Table~5.1 in Ref.~\cite{Schlippert14}.}
\label{tab:systematics}
\end{table}

The statistical uncertainty~$\sigma_\eta=\stat$ after $\SI{4096}{s}$ of integration time is dominated by technical noise in the potassium interferometer. This limitation can be improved in several ways. 

For instance, the implementation of a sequence preparing a single $m_F$ state would reduce the number of background atoms that currently lower the contrast. Moreover, selecting a narrower velocity class, as well as lower temperatures to begin with, would improve the beam splitting efficiency and consequently the contrast, too.

\section{Atom-chip based BEC interferometry}
\label{sec:BEC_interferometry}

Today's generation of inertial sensors based on atom optics typically operates with cold atoms, released or launched from an optical molasses exemplified by the experiment discussed in the previous section. The velocity distribution and the finite size of these sources limit 
the efficiency of the beam splitters as well as the analysis of the systematic uncertainties. 
We can overcome these limitations by employing ensembles with a momentum distribution well below the photon recoil limit, which can be achieved with BECs. 

After reaching the regime of ballistic expansion, where the mean-field energy has been converted into kinetic energy, the momentum distribution of a BEC can be narrowed down even further by the technique of delta-kick collimation~(DKC)~\cite{Chu86OL,Ammann97PRL,Morinaga99PRL,Muentinga13PRL,Kovachy15PRL}. Moreover, atom-chip technology offers the possibility 
to generate a BEC and perform DKC in a fast and reliable fashion, paving the way for miniaturized atomic devices. 
We devote Sec.~\ref{sec:DKC} to an introduction to DKC and note that the results reported in this section have originally been published 
in Ref.~\cite{Muentinga13PRL}.

The use of BECs allows us to implement Bragg and double Bragg diffraction with efficiencies
above 95\%, and thus to perform interferometry with high contrast. 
In Sec.~\ref{sec:Tiltmeter} we demonstrate a quantum tiltmeter using a MZI based on double Bragg diffraction with a tilt precision of up to $4.5\,\upmu$rad. The results presented in this section have originally been published in Ref.~\cite{Ahlers16PRL}. 

In Sec.~\ref{sec:Compact_baseline} we discuss an experiment combining double Bragg diffraction and Bloch oscillations, 
where we have implemented \cite{Abend16PRL} a relaunch procedure with more than 75\% efficiency in a retro-reflected optical lattice. 
We emphasize that here we rely on a {\it single} laser beam only, which also comprises the beam splitter resulting in a set-up of significantly reduced complexity. 

Our relaunch technique allowed us to build a gravimeter on a small baseline with a comparably large interferometry time of~$2T=50$\,ms 
in the MZI for a fixed free-fall distance. At a high contrast of $C=0.8$ the interferometer reaches an intrinsic sensitivity to gravity of 
\begin{equation}
\Delta g/g=1.4\cdot 10^{-7}\,.
\end{equation}
A key element of this result was the state preparation comprising DKC and Stern-Gerlach-type deflection, 
which improved the contrast and reduced the detection noise. The results presented in this section have originally been reported in Ref.~\cite{Abend16PRL}.

\subsection{Delta-kick collimation}
\label{sec:DKC}

The desire to reach long expansion times serves as a motivation for DKC by a magnetic lens~\cite{Smith08JPB}. 
Recent experiments have shown expansion rates corresponding to a few nK in 3D in the drop tower~\cite{Muentinga13PRL} with QUANTUS-1, or even pK in 2D in a 10\,m-fountain~\cite{Kovachy15PRL}. These widths in momentum space are smaller than those of 
the coldest reported condensates~\cite{Leanhardt03Science}. For a detailed study of DKC using the QUANTUS-1 apparatus in the drop tower and also 
on ground we refer to Refs.~\cite{Wenzlawski13,Krutzig14}.

After the release from the trap, the BEC starts to expand freely and falls away from the trap due to the gravitational acceleration. 
During the first milliseconds after release most of the mean-field energy is converted into kinetic energy. 
The required time for this conversion depends on the atomic density of the condensate, and hence, on the steepness of the trap 
from which the BEC is released. The final expansion rates in the ballistic regime can be precisely evaluated by time-of-flight measurements. 

Even a BEC has a non-zero velocity spread leading after some expansion time to an increased cloud size, and possibly 
to a reduction in performance of an atom interferometer due to an increased detection noise, or larger contributions to systematic uncertainties. 
A condensate released from a shallow trap expands slowly enough to perform experiments with short free-fall times. Moreover, these condensates have a momentum width that is small enough to reach high Bragg diffraction efficiencies. 

Indeed, for trapping frequencies of $f_{(x,y,z)}=(18,\,46,\,31)$\,Hz and $N=10^4$ atoms the resulting expansion rate along 
the beam-splitting axis is $\sigma_v\approx\, 750\,\upmu\mathrm{m}/\mathrm{s}\approx\, 0.125\,\hbar k/m$, 
which corresponds to an effective temperature in the beam direction of about 5 to 10\,nK. 
Thus, for this number of atoms the mean-field conversion in a time of $t_{\mathrm{exp}}\gtrsim 10$\,ms is acceptable for interferometry.

\begin{figure}[h]
\centering
\includegraphics[width=.7\linewidth]{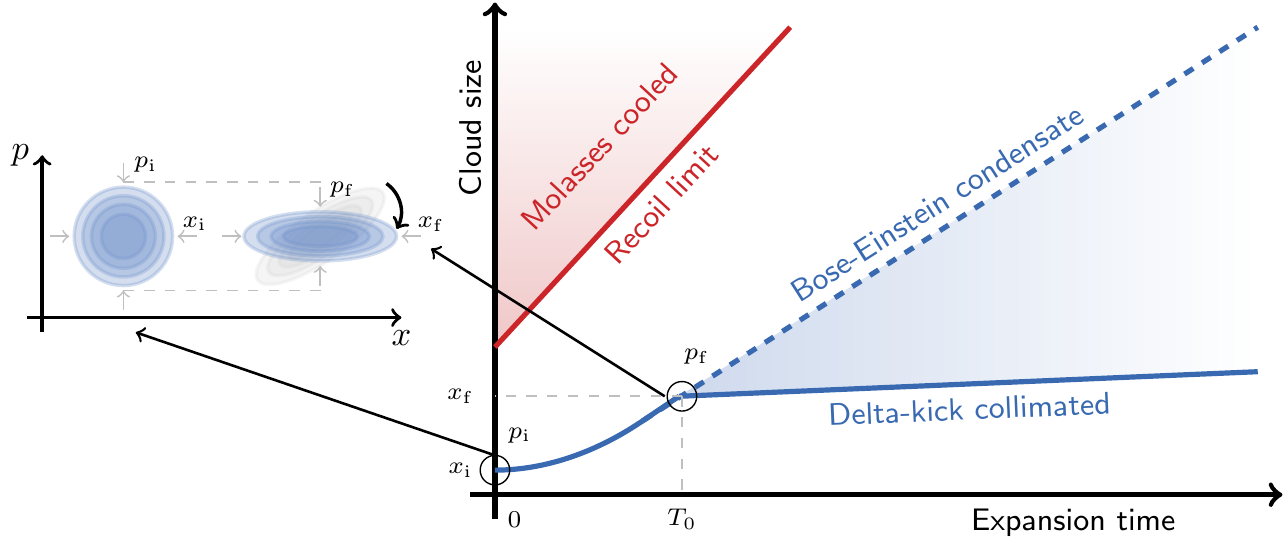}
\caption{Principle of delta-kick collimation (DKC) explained in phase space. After release an ensemble of cold atoms has an initial distribution in the phase space 
spanned by position $x$ and momentum $p$. 
After a time $T_0$ the cloud has expanded in space giving rise to a tilted ellipse which is then rotated in phase space by a collimating pulse. 
This figure is an adaptation of Fig.~1.3(c) in Ref.~\cite{Abend17}.}
\label{fig:dkc}
\end{figure}

However, for larger densities of the condensate, the use of DKC becomes necessary. In Fig.~\ref{fig:dkc} we illustrate the essential idea of DKC with the time evolution of a phase-space distribution. 
The shearing in phase space caused by the free evolution of the condensate leads to a tilted ellipse that can be rotated by applying a harmonic potential 
for a suitable time~$\tau_{\mathrm{DKC}}$, so as to align its major axis with the $x$-axis.

Indeed, when we consider the potential~$V_{\mathrm{DKC}}(x)\equiv m \omega_{\mathrm{DKC}}^2x^2/2$ 
applied for a short time~$\tau_{\mathrm{DKC}}$, the resulting change~$\Delta p$ in the momentum~$p$ is approximately given by 
\begin{equation}
\Delta p \cong F(x) \tau_{\mathrm{DKC}} = -\frac{\textrm{d}V_{\textrm{DKC}}}{\textrm{d} x} \tau_{\mathrm{DKC}} = -m\omega_{\mathrm{DKC}}^2x\,\tau_{\mathrm{DKC}}\,.
\end{equation}

Assuming that most of the expansion takes place in the ballistic regime, and that the spatial size~$x_{\mathrm{f}}$ 
when the DKC pulse is applied at time~$T_0$ is much larger than the initial size~$x_{\mathrm{i}}$, 
such that $x_{\mathrm{f}}\approx p_{\mathrm{i}}T_0/m$, the induced momentum change reads
\begin{equation}
\Delta p=-p_{\mathrm{i}}  \omega_{\mathrm{DKC}}^2    \tau_{\mathrm{DKC}}T_0\,.
\end{equation}
The initial momentum width $p_{\mathrm{i}}$ considered here includes the contribution from the mean-field energy that was converted into kinetic energy. 

In complete analogy to the power $P$ of an optical lens the strength of a delta kick is defined by $S\equiv\omega_{\mathrm{DKC}}^2\tau_{\mathrm{DKC}}$. For a given $T_0$ the optimal choice that leads to $\Delta p=-p_{\mathrm{i}}$ corresponds to $S=1/T_0$, or equivalently, to
\begin{equation}
\label{eq: dkc_optimal_choice}
\omega_{\mathrm{DKC}}^2\tau_{\mathrm{DKC}} T_0=1\,.
\end{equation}

However, even in this case the minimum value attainable for the final momentum width is limited by Liouville's theorem, that is 
conservation of phase-space volume. This minimal value is determined by the ratio of the phase-space volume of the initial ensemble, and the spatial width $x_{\mathrm{f}}$ 
when the DKC pulse is applied. In principle this limitation can be reduced by increasing $T_0$ which leads to a larger $x_{\mathrm{f}}$.

In a realistic implementation uncertainties in the relevant parameters lead to an uncertainty $\delta p$ in the induced momentum change 
which should be smaller than the targeted final momentum width $p_{\mathrm{f}}$, namely
\begin{equation}
\label{eq: dkc_momentum_requirement}
\delta p=p_{\mathrm{i}} \delta(\omega^2_{\mathrm{DKC}}\tau_{\mathrm{DKC}}T_0)< p_{\mathrm{f}}\,.
\end{equation}

When we take into account the optimal choice, Eq.~\eqref{eq: dkc_optimal_choice}, the requirement, Eq.~\eqref{eq: dkc_momentum_requirement}, can be expressed in terms of the relative errors for the individual parameters, that is
\begin{equation}
2\frac{\delta\omega_{\mathrm{DKC}}}{\omega_{\mathrm{DKC}}} + \frac{\delta\tau_{\mathrm{DKC}}}{\tau_{\mathrm{DKC}}} + \frac{\delta T_0}{T_0} < \frac{p_{\mathrm{f}}}{p_{\mathrm{i}}}\,.
\end{equation}

For ground-based experiments, the waiting time~$T_0$ and other quantities determining the errors are not only limited by technical means, 
but also by the free fall away from the chip, which reduces the trap frequencies of the potential, and restricts~$T_0$ to times smaller than~$6$\,ms. 
For a shallow trap the expansion after this time is not even in the ballistic regime, and the mean-field potential is still non-negligible. 
Thus, a release from the trap with a faster initial expansion is required to increase~$x_{\mathrm{f}}$ prior to collimation. 

We conclude our brief review of DKC by mentioning that the anharmonicities of the generated potential represent another source of errors. Indeed, they cause deformations when the condensates are too large.

\subsection{Quantum tiltmeter based on double Bragg diffraction}
\label{sec:Tiltmeter}

An interesting extension of Bragg diffraction occurs when a retro-reflected pair of laser beams interacts with atoms that are at rest 
with respect to the mirror, such that the Doppler shift~$\omega_{\mathrm{D}}$ vanishes. 
In this case the transitions with opposing effective wave vectors are degenerate in frequency, and simultaneously diffract 
the atomic wave packets in both directions. The difference in momenta between both arms in an interferometer is then increased to $4\hbar k$. 
This symmetric diffraction called \quot{double Bragg diffraction} was proposed in Ref.~\cite{Giese13PRA} as a generalization 
of Bragg diffraction in complete analogy to double Raman diffraction~\cite{Leveque09PRL}, and experimentally demonstrated in Ref.~\cite{Ahlers16PRL}. 
We emphasize that the traditional Bragg diffraction can be referred to as single or uni-directional Bragg diffraction, 
since only a single pair of laser beams drives the transition, while the other one is off resonant. 

\subsubsection{Rabi oscillations}

\begin{figure}[!h]
\includegraphics[width=\linewidth]{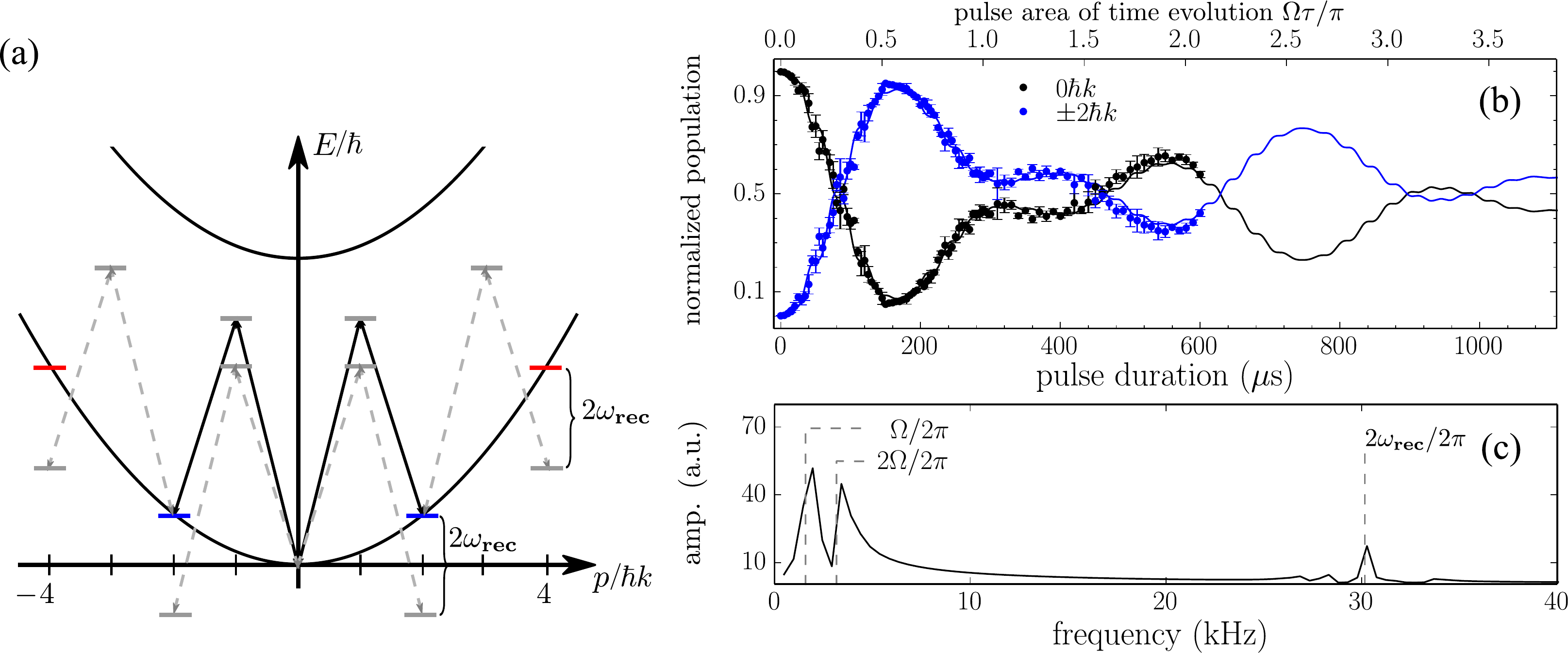}
\caption{First-order double Bragg diffraction represented by the corresponding transitions~(a), the comparison~(b) between experimental observations and numerical simulations of the Rabi oscillations in the normalized populations 
as a function of the square-pulse duration, and their spectral decomposition~(c). 
The energy diagram in (a) shows the resonant (solid lines) and off-resonant (dashed lines) 
light-induced transitions between the atomic momentum states $\left|0 \hbar k\right\rangle$ (black), 
$\left|\pm 2 \hbar k\right\rangle$ (blue) and $\left|\pm 4 \hbar k\right\rangle$ (red). 
The experimental values (squares) and numerical simulations (solid curves) based on Ref.~\cite{Giese13PRA} displayed in (b) are in good agreement. 
The frequency spectrum (c) of the simulated population $N_1/N_\text{tot}$ displays components close to $2 \omega_\mathrm{rec}$ 
which stem from the off-resonant couplings depicted in (a) by dashed lines. 
The broad double-peaked structure at the Rabi, and twice the Rabi frequency, which is a consequence of the detuned three-level system 
with non-vanishing $p_0$ and/or $\delta p$, leads to the modulation of the oscillations observed in (b). 
This figure is reproduced from~\cite{Ahlers16PRL} with permission of the authors, copyright American Physical Society (2016).}
\label{fig:ddiff_rabi}
\end{figure}

In Fig.~\ref{fig:ddiff_rabi}(a) we show the coupling scheme for first-order double Bragg diffraction. 
The transition frequency is given by the Bragg condition, Eq.~\eqref{eq:braggcondition}, with the recoil frequency~$\omega_{\mathrm{rec}}$. 

Due to the intrinsic symmetry of the diffraction process, a beam splitter of this kind offers several advantages for atom interferometry. Most prominently, the populations of the output ports no longer depend on the laser phase~$\Delta\phi^{\mathrm{laser}}$,  since the wave functions of the center-of-mass motion in the two arms 
have the same laser phase imprinted during each pulse\footnote[4]{Needless to say, this feature is only present when we neglect off-resonant processes.}.
Moreover, we can choose the order of the diffraction process by matching the detuning $\delta$ with the kinetic energy gained during the scattering.

Rabi oscillations for double Bragg diffraction are more complicated than for single Bragg diffraction, 
since more states and transitions have to be taken into account. 
In the case of first-order double Bragg diffraction $\delta$ is chosen to correspond 
to the recoil frequency $\omega_\text{rec}\equiv (2\hbar k)^2/(2m \hbar)$ inducing resonant transitions between the momentum states $\left|0\right\rangle$ and $\left|\pm 2 \hbar k\right\rangle$ depicted in Fig.~\ref{fig:ddiff_rabi}(a) by solid lines, together with off-resonant transitions to these, and higher momentum states indicated by dashed lines. 
Being off-resonant, the latter transitions are substantially suppressed.

When the width of the momentum distribution is small enough we can observe Rabi-type oscillations~\cite{Ahlers16PRL} as a function of the atom-light interaction time as shown in Fig.~\ref{fig:ddiff_rabi}(b). Here we also display numerical simulations \cite{Giese13PRA} 
of the Rabi oscillations and the normalized atom populations are defined as
\begin{equation}
\left.
\begin{aligned}
n_0 \equiv N_0/N_\text{tot} \quad \text{and} \quad
&n_j \equiv (N_{-j} + N_j)/N_\text{tot}\quad (j=1,2)\,,
\end{aligned}
\right.
\label{populations}
\end{equation}
where $N_j$ is the number of atoms in the momentum state~$\ket{2j\hbar k}$  with $j=-2,-1,0,1$ and $2$, and $N_\text{tot} \equiv \sum_{j = -2}^2 N_j$.

\subsubsection{Tilt measurements}
In most atom interferometers tilt variations can be significant and lead to an uncertainty in the quantity being measured. 
Hence, these devices are either designed to minimize the effect of tilts or do not allow to distinguish a tilt from other sources inducing a phase shift~\cite{Altin13NJP}.

\begin{figure}[!h]
\includegraphics[width=\linewidth]{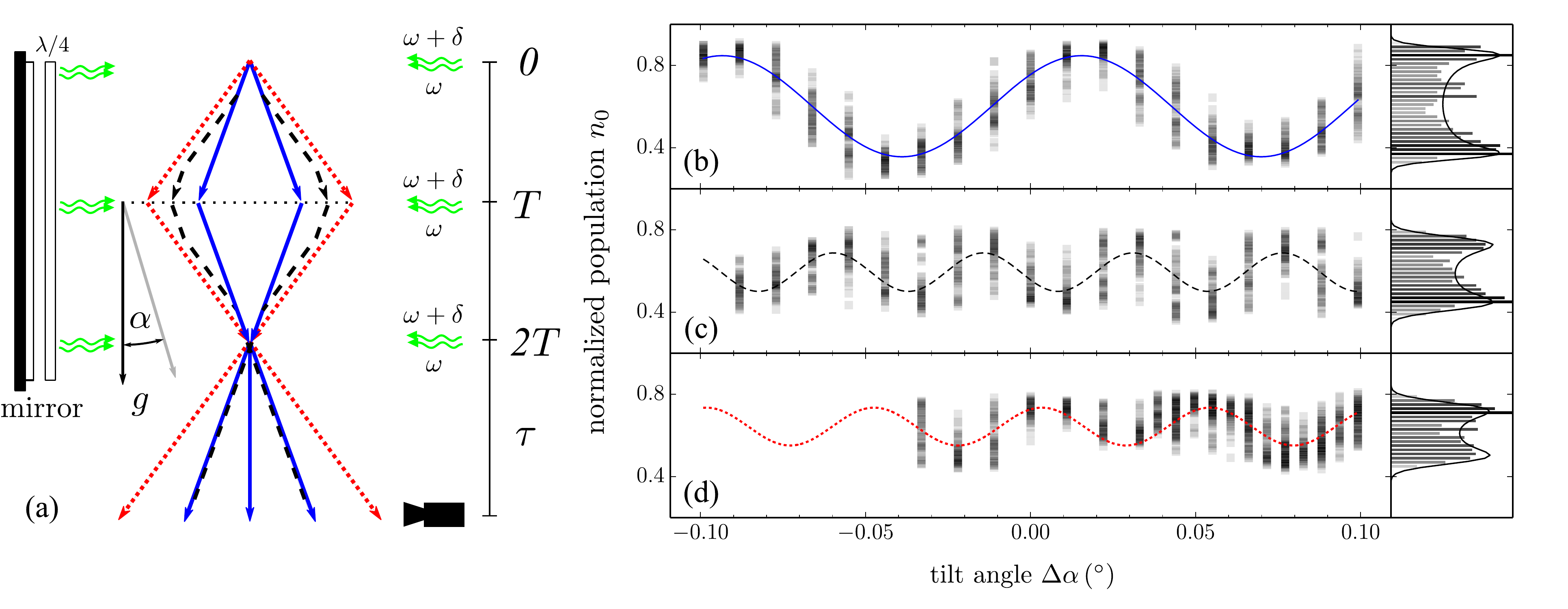}
\caption{Three double Bragg interferometers employed as tiltmeters (a) and the interference signals corresponding to 
first-order~(b), successive first-order~(c), and second-order Gaussian pulses~(d) as a function of the tilt angle $\Delta \alpha$. 
For each angular step in $\Delta \alpha$, the normalized population~$n_0$ in the exit ports is measured 50 times. 
The blue solid, black dashed and red dotted lines represent sinusoidal fits of those datasets. 
Fitting the histograms of~$n_0$ over a range of tilt settings corresponding to one, 
or two complete fringe periods, and fitting them with a theoretical distribution (black) which assumes that all noise sources combined 
with the tilt scan lead to an approximately uniform phase-shift distribution, yield contrasts of 43\%, 29\% and 23\%, respectively. 
Further analysis reveals a tilt precision of $4.5\,\upmu$rad, $5.9\,\upmu$rad, and $4.6\,\upmu$rad, respectively. 
This figure is reproduced from Ref.~\cite{Ahlers16PRL} with permission of the authors, copyright American Physical Society (2016).}
\label{fig:ddiff_scheme_fringes}
\end{figure}

We have designed an atom interferometer to measure slight deviations from the horizontal direction with respect to gravity.
In this quantum tiltmeter we diffract a delta-kick collimated BEC with small initial momentum and low expansion rate off laser beams and induce first- or higher-order double Bragg transitions. 

Figure~\ref{fig:ddiff_scheme_fringes}(a) illustrates the corresponding symmetric geometry emerging from a first-order (blue solid lines) double Bragg process, where the initial wave packet is split, redirected and recombined. A stepwise tilt of the whole apparatus changes the orientation $\alpha$ of the interferometer with respect to gravity $g$ and induces a phase shift. As a result, the interference signal, that is the normalized populations at the exit ports, exhibits oscillations as a function of the change $\Delta \alpha$ in the tilt angle exemplified by Fig.~\ref{fig:ddiff_scheme_fringes}(b). 
This process can be extended to  successive first-order (black dashed lines) and second-order (red dotted lines) double Bragg processes 
as displayed in Fig.~\ref{fig:ddiff_scheme_fringes}(c) and (d).

\subsection{Sensitive atom-chip gravimeter on a compact baseline}
\label{sec:Compact_baseline}

Quantum sensors for gravimetry based on cold atoms have been with us for more than two decades \cite{Kasevich92APB,Peters99Nature,Hu13PRA}, and 
can reach today accuracies competitive with falling corner cube gravimeters~\cite{Merlet10Metrologia}. 
However, {\it compact} gravimeters using BECs have been demonstrated only recently~\cite{Debs11PRA,Abend16PRL}.

Due to their small extent and expansion rates, BECs which are delta-kick collimated have attracted attention for large-scale devices 
on ground~\cite{Dickerson13PRL}, and in space missions~\cite{Aguilera14CQG}. In light of these experiments systematic uncertainties specific to 
BECs have been analyzed~\cite{Schubert13arxiv, Hartwig15NJP}, and novel techniques have been introduced~\cite{Sugarbaker13PRL}. 
These achievements will allow sensors relying on BECs to target sub-$\upmu$Gal accuracies in the near future, and to overcome 
current limitations set by cold atoms~\cite{Hu13PRA,Bidel13APL,Freier16CS,Fang16CS}. 

The use of an atom chip for all preparation steps, and as a retro-reflector is the novelty of our approach \cite{Abend16PRL} summarized in this section. 
Although our experiment is a proof-of-principle, it nevertheless represents an important pathway to the application of an atom-chip gravimeter 
for precision measurements important for example in geodesy.

\subsubsection{Relaunch of atoms in a retro-reflected optical lattice}
In order to increase the observation time of a BEC on a small baseline, we have developed a simple but extremely valuable method to relaunch a BEC using Bloch oscillations in a retroreflected or dual lattice. Our procedure differs substantially from previous ones, which either rely on (i) two crossed beams reflected from a mirror 
surface~\cite{Sugarbaker14PHD}, (ii) two opposing beams~\cite{Andia13PRA}, 
(iii) velocity selection from a molasses~\cite{Charriere12PRA,Altin13NJP,Mazzoni15PRA}, or (iv) the transfer of only a few photons 
from a standing wave~\cite{Impens06APB,Hughes09PRL}. Indeed, we employ a retro-reflected optical lattice which is a common configuration in atomic sensors. 

The novelty, and at the same time the challenge of our method is that for the experiment we use only a {\it single} beam along the vertical axis with linearly polarized laser light of two co-propagating frequency components. They form in total four lattices: 
two moving lattices with opposite velocity, and two additional ones at rest. 

We perform the relaunch procedure in three steps to avoid a zero-crossing of the velocities of the lattices:
(i) a lattice deceleration, (ii) a momentum inversion pulse, and (iii) a lattice acceleration. 

Our sequence starts by loading the atoms adiabatically into one of the lattices where the Bloch oscillations  for deceleration are performed 
until the atomic motion is almost stopped. The atoms are then adiabatically unloaded from the lattice with only a few~$\hbar k/m$ of residual velocity. 
	
After a small waiting time to carefully match the resonance condition to the velocity of the atoms, a higher-order double Bragg diffraction pulse is applied which inverts the momentum.
	
Finally, a second lattice acceleration sequence follows, which precisely speeds up the atomic ensemble to launch it on a parabolic trajectory with an adjustable apex close to the atom-chip surface.

\begin{figure}[h]
\includegraphics[width=\linewidth]{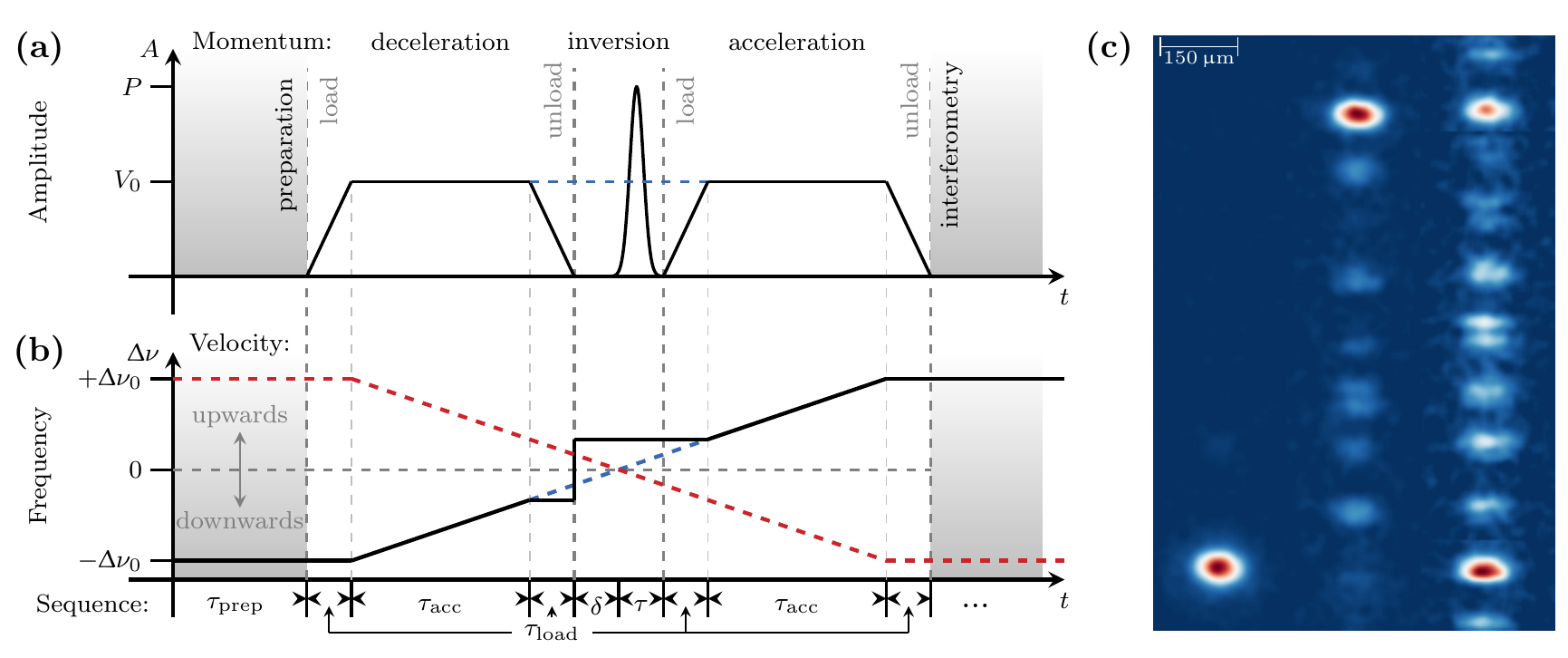}
\caption{Relaunch of atoms in a retro-reflected optical lattice represented by an amplitude (a) and frequency modulation (b) of the lattice and compared to a single-lattice acceleration together with density plots (c) of the final momentum distribution. The relaunch sequence circumvents the zero-velocity crossing of the dual lattice 
by a double Bragg pulse inverting the momentum. The two frequency components of the dual lattice are depicted in red and blue. In (c) we compare the momentum inversion with a 16\,$\hbar k$ 
double Bragg diffraction pulse (middle) to the one after the deceleration (left), and a simple sweep through resonance 
with Bloch oscillations (right). This figure is an adaptation of Figs.~5.17 and~5.25(b) in Ref.~\cite{Abend17}.}
\label{fig:relaunch_sequence}
\end{figure}

In Fig.~\ref{fig:relaunch_sequence}(a) and (b) we show the amplitude and frequency modulation of the dual-lattice light used to perform 
the relaunch sequence. Compared to a single-lattice acceleration this scheme is rather complex, 
since the additional lattices from the retro-reflection are shifted out of resonance by the Doppler effect of the falling atoms. Unfortunately, this effect does not allow the acceleration of the lattices in a way that the atomic ensemble crosses the zero-momentum state 
without losing a major fraction of atoms as depicted in Fig.~\ref{fig:relaunch_sequence}(b). 

These losses result from the fact that when the atoms are at rest, 
the two moving optical lattices are both on resonance with the atoms --- one attempting to move the atoms upwards, and the other one to move them downwards. 
This feature reduces the fraction of atoms that are launched upwards to about one-half of the atom number achieved by double Bragg diffraction. 
Moreover, when the velocity of the atoms almost vanishes, non-adiabatic transitions arise due to parasitic acceleration in the non-resonant lattice. They remove atoms from the upward moving lattice, and further reduce the number of launched atoms to about one quarter. 

Fortunately, a combination of Bloch oscillations in an optical lattice together with higher-order Bragg diffraction prevents these losses. 
In this scheme most of the momentum is transferred via Bloch oscillations with an efficiency close to unity to stop and launch the atoms. Since only a smaller fraction of momentum needs to be transferred by a single Bragg pulse, this sequence maintains a high overall efficiency.

\subsubsection{Experimental sequence of the atom-chip gravimeter}
\begin{figure}[h]
\includegraphics[width=\linewidth]{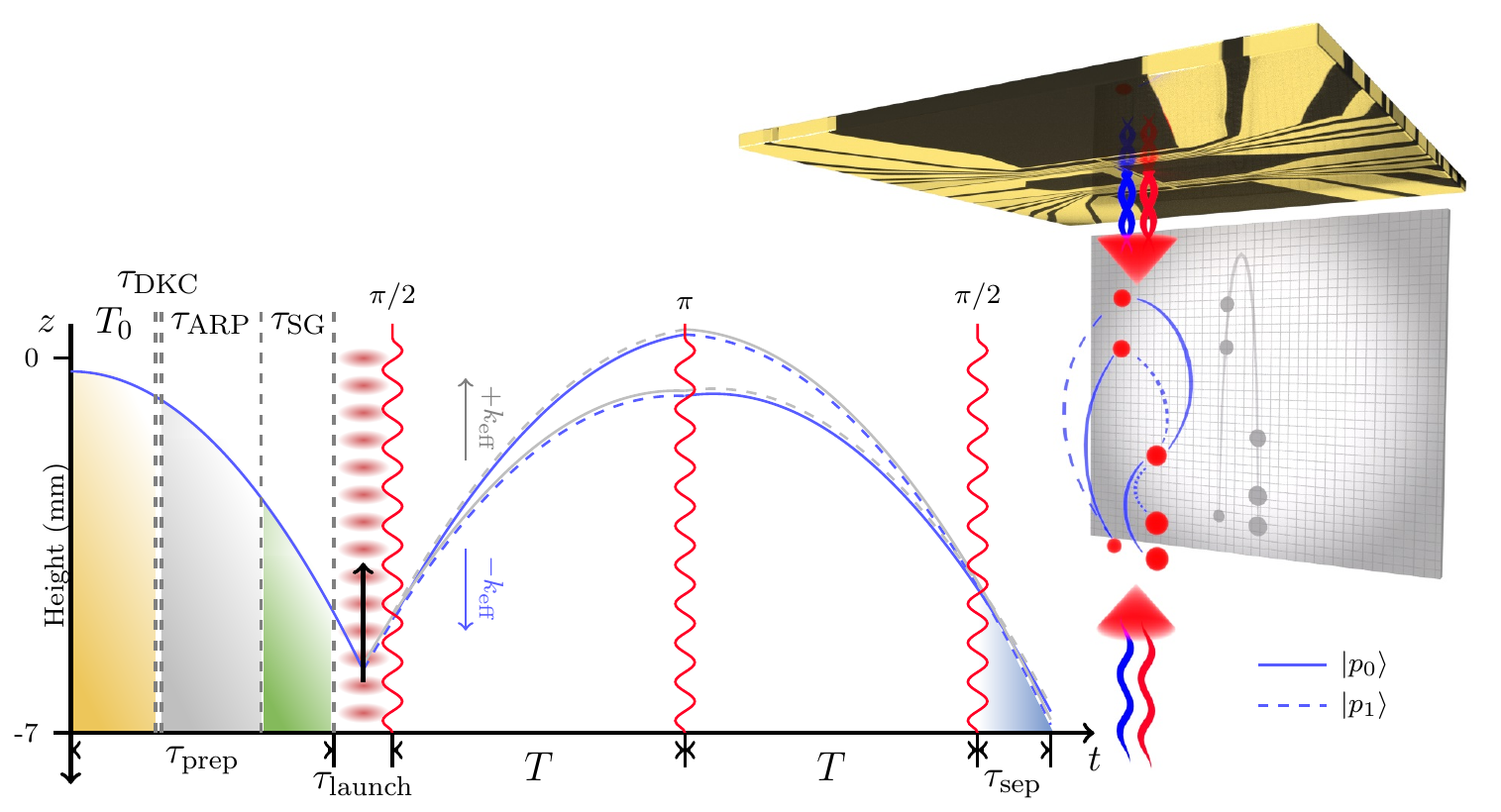}
\caption{Atom-chip gravimeter with an extended free-fall time of up to $\tau_{\mathrm{ToF}}\approx100$\,ms represented 
in spacetime (left) and space (right). The preparation of the atomic ensemble is performed during $\tau_{\mathrm{prep}}$ 
before the relaunch. The elongated free-fall time allows us, after the expansion time $T_0$, to employ DKC (for the time $\tau_{\textrm{DKC}}$) as well as adiabatic rapid passage (for the time $\tau_{\textrm{ARP}}$) and Stern-Gerlach-type 
deflection (for the time $\tau_{\textrm{SG}}$) to reduce the expansion rate, and to remove atoms remaining in magnetic sensitive states. 
The relaunch in the retro-reflected optical lattice is realized as described. 
The MZI features up to third-order Bragg diffraction, pulse separation times up to $T=25\,$ms, and a detection 
after a separation time of~$\tau_{\mathrm{sep}}\ge10$\,ms. This figure is reproduced from Fig.~6.3 in Ref.~\cite{Abend17}.}
\label{fig:fountaingravi_sequence}
\end{figure}

Bragg diffraction combined with the relaunch allows us to implement a sensitive gravimeter with the atom chip. 
Figure~\ref{fig:fountaingravi_sequence} shows the spacetime diagram of the fountain geometry.

Subsequent to the adiabatic rapid passage we also perform Stern-Gerlach-type deflection by a magnetic field pulse with a duration of~$\tau_{\mathrm{SG}}=7$\,ms using the Z-wire on the chip. 
In this way we remove the atoms remaining in magnetic sensitive states leading to an enhanced contrast. 

The maximum value of the preparation time~$\tau_{\mathrm{prep}}$ is limited to 34\,ms 
due to the end of the detection region $7$\,mm below the chip. 
The relaunch process has an overall duration of $\tau_{\mathrm{launch}}=2.9$\,ms. 
With a relaunch realized after the largest waiting time of~$\tau_{\mathrm{prep}}=33.2$\,ms the total time of flight~$\tau_\mathrm{ToF}$ after initial release of the atoms is greatly increased to $\tau_\mathrm{ToF}=97.6$\,ms. 

The interferometer sequences start after the atoms have been launched on their fountain trajectories. 
The final waiting time~$\tau_{\mathrm{sep}}$ after~$\tau_\mathrm{ToF}>90$\,ms to separate the output ports can be reduced 
from~$\tau_{\mathrm{sep}}\ge20$\,ms to~$\tau_{\mathrm{sep}}\ge10$\,ms provided DKC is used. 
The remaining time~$2T\equiv\tau_{\mathrm{ToF}}-\tau_{\mathrm{prep}}-\tau_{\mathrm{launch}}-\tau_{\mathrm{sep}}<51$\,ms can be entirely used 
for the interferometry, which allows us to use a pulse separation time as large as $T=25$\,ms. 
The limit in $T$ on our current baseline of 7\,mm is reached with $T=25$\,ms. Any further extension beyond this value
would result in a reduced contrast due to insufficient port separation.

State-of-the-art Raman-type gravimeters routinely operate with pulse separation times of 70\,ms~\cite{Fang16CS} 
or larger~\cite{Hu13PRA,Freier16CS}. To further increase the scale factor, not only first-order but also higher-order Bragg diffraction 
can be implemented in the MZI. At the moment third-order Bragg diffraction has an efficiency of above $90\%$. 
With future improvements, already fourth-order Bragg diffraction will compensate a decrease by a factor 
of two in the pulse separation time~$T$.

To perform higher-order Bragg diffraction, we have used shorter beam-splitter pulses, but at larger laser powers. 
More specifically, we have employed Gaussian-shaped pulses of widths~$\sigma_{\tau}=12.5\,\upmu$s. 

The first $\pi/2$-pulse of the MZI follows with a short delay of one millisecond after the relaunch 
to maximize the time available for interrogation. To avoid $\pi$-pulses at the apex the timing of the MZI needs to be placed {\it asymmetrically} around the apex of the fountain. In this case, both lattices are on resonance, and losses 
due to double Bragg diffraction, and standing waves disturb the $\pi$-pulse. 
As a consequence, the Doppler detuning $\delta$ of the $\pi$-pulse should satisfy the condition $\delta > 100$\,kHz, 
or correspondingly, the time difference to the apex should be of the order of $7-8$\,ms. 
Consequently, the separation time of the outputs is always larger than~$\tsep >14$\,ms.

A larger momentum transfer slightly reduces the free-fall time, because depending on the direction 
of the momentum transfer~$\pm \hbar \keff$ the atoms are either kicked towards the atom chip, 
or downwards such that they leave the detection region faster. By choosing the momentum transfer 
of the second and final lattice acceleration in the relaunch sequence according to the direction of 
the momentum transfer of the beam splitter, the height of the parabola can be maximized. 
As a consequence, the pulse separation time~$T$ can be held constant, independently of $\pm \hbar \keff$, 
as depicted in Fig.~\ref{fig:fountaingravi_sequence}. In both cases the free fall time~$\tau_{\mathrm{ToF}}$ of the atoms is slightly reduced due to the recoil imprinted during the beam splitting process.

\subsubsection{Analysis of the interferometer output}
\begin{figure}[h]
\includegraphics[width=0.8\linewidth]{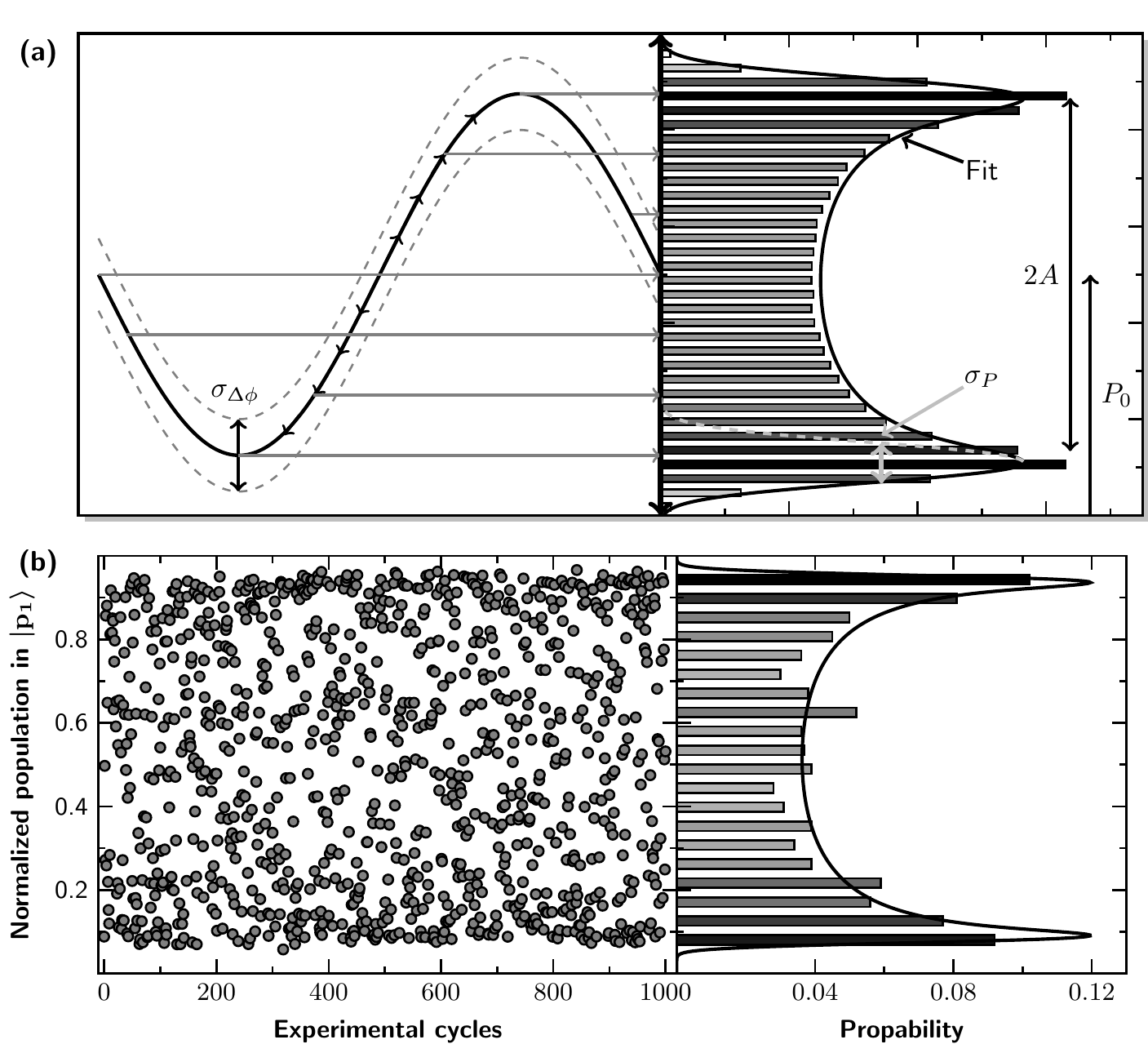}
\caption{Extraction of the interferometer phase from a noisy signal with the help of the probability density. 
Vibrational noise completely washes out a sinusoidal signal~(a), and only fluctuations are left to be measured 
in the normalized population between the two output ports. The corresponding histogram exhibits 
a characteristic double-peak structure from which we extract the contrast~$C\equiv A/P_0$ given by the amplitude~$A$ of the sinusoidal signal and its mean $P_0$. The output (b) of a MZI in a fountain geometry 
using delta-kick collimated ensembles with a pulse separation time of $T=25$\,ms and  
first-order Bragg diffraction resembles noise after roughly 1\,000 measurements have been taken. 
Nevertheless, we can obtain a contrast of $C=0.86$ and a technical noise~$\sigma_{\Delta\phi}$ close to the shot noise from the histogram. This figure is an adaptation of Figs.~6.11 and~6.17(c) in Ref.~\cite{Abend17}.}
\label{fig:fountaingravi_results}
\end{figure}

Since for pulse separation times larger than~$T>5$\,ms the background vibration level
leads to a complete loss of the fringe pattern~\cite{Mcdonald13PRA} we can no longer employ the common fringe fit method, or an Allan deviation analysis. 
Even when the laser phase and the chirp rate are identical in subsequent measurements, 
the output phase scatters over multiple $2\pi$-intervals, as illustrated in Fig.~\ref{fig:fountaingravi_results}(a). 
As a consequence, at these high levels of sensitivity the readout of a gravity-induced phase shift is impossible without having 
additional information about the vibrations during the interferometry, such as seismic correlation.

We emphasize that the beam splitters still operate with high fidelity, and oscillations between the output ports are clearly visible. 
However, the simple peak-to-valley, or standard-deviation calculation over- or underestimates the contrast, and does not yield 
a useful noise analysis for the output signals. 

One method to solve the problem of distinguishing between useful contrast~$C$, 
and technical noise~$\sigma_{\Delta\phi}\equiv\sigma_P/C$, rests on a histogram analysis revealing the contrast of an interferometer 
without relying on fringe visibility~\cite{Geiger11NatureComm}. Here $\sigma_P$ denotes the fluctuation in the measured population. 

In this approach the output signals of a data set are first split into bins of equal size containing the normalized population~$P$, 
and then the data points within each interval are counted. The resulting histogram shows a characteristic double-peak structure 
reflecting the sinusoidal dependence of the interference signal. This structure results from a simple noise model, 
assuming that for a completely random signal the probability to find an output state with a normalized population 
at top or bottom of a sinusoidal fringe pattern is larger than at the middle. 
We emphasize that this method requires sufficient statistics over several hundred experimental cycles with a stable signal.

We extract the contrast~$C\equiv A/P_0$ from a fit of the distribution according to Fig.~\ref{fig:fountaingravi_results}(b), with the amplitude~$A$ of the signal and its mean $P_0$. As input for the fitting routine a kernel density estimation (KDE) 
of the data points is used, rather than the histogram 
itself\footnote[5]{This evaluation yields a slightly better intrinsic sensitivity compared to 
the value published in Ref.~\cite{Abend16PRL}.}. 

In a KDE each data point is weighted with a Gaussian function of a fixed width~$\sigma_{\mathrm{KDE}}=0.01$. 
All Gaussian functions are then added and the signal normalized such that in the end a continuous distribution 
properly reflects the density of data points without loss of information. 

For the width~$\sigma_{\mathrm{kde}}$ it is only of importance to choose a value smaller than the expected technical noise. 
If the histogram is fitted for a small number~$n_{\mathrm{bin}}$ of bins, the histogram would yield insufficient information 
to be properly fit, and would lead to a larger uncertainty in the extracted parameters.

In our fountain geometry the expected gain in contrast is between 5\% and 10\% and results from two factors: (i) We can employ delta-kick collimated BECs with ultra-slow expansion and (ii) the Stern-Gerlach-type deflection purifies the magnetic sub-states. With this configuration the pulse separation time can even be extended to $T=25$\,ms. 
At this time, the output ports are still separated due to the smaller final size of the clouds. 

The measurements and evaluation depicted in Fig.~\ref{fig:fountaingravi_results}(b) are for our MZIs formed by 
first-order Bragg diffraction and $T=25$\,ms. The histogram analysis reveals that the contrast remains at $C=0.86$. 
The technical noise level~$\sigma_{\Delta\phi}=14$\,mrad is extracted from the widths of the outer peaks in the fit 
to the density distribution, which is close to the calculated shot noise of~$11$\,mrad for $N=8\,000$ atoms. 
This remarkable result is due to the interplay between the high-fidelity Bragg diffraction and the DKC. 

These ultra-slow expansion rates allow for even longer flight times, and also give rise to a boost in sensitivity. Indeed, the largest intrinsic sensitivity~$\Delta g/g = 1.4\cdot 10^{-7}$ was observed 
after a $\tau_{\mathrm{ToF}}=97.6$\,ms at a noise of $\sigma_{\Delta\phi}=14$\,mrad. 
This achievement represents an important step towards compact but precise sensors.

\section{Outlook}
\label{sec:Outlook}

Atom interferometry is a cornerstone of precision measurements with a wealth of promising applications. In particular, we expect atomic gravimeters based on BECs to reach sub-$\upmu$Gal accuracies in the near future. 
In Sec.~\ref{sec:systematics} we highlight the origins of measurement uncertainties, and present mitigation strategies 
for future devices.

The tools and methods for BEC interferometry outlined in these lecture notes have opened the path towards significantly enhanced scale factors due to an extended free evolution time.
For this reason we focus in Sec.~\ref{sec:VLBAI} on very long baseline atom interferometers. Moreover, we devote Sec.~\ref{sec:space} 
to space-borne devices and analysize the potential for future gravity measurements as well as tests of fundamental physics 
such as the UFF.

\subsection{Reduced systematic uncertainties in future devices}
\label{sec:systematics}

The main drive for sensors based on BECs is the reduction of systematic uncertainties. We now assess the potential of an atom chip gravimeter to reach sub-$\upmu$Gal accuracies. For this purpose we identify in Table~\ref{tab:futuresystematics} the origins of the largest contributions to the measurement uncertainty and suggest mitigation strategies.

\begin{table}[h]
\centering
\small\renewcommand{\arraystretch}{1.4}
	
\begin{tabular}{| p{2.8cm} | p{5.45cm} | c | c|}
\hline
Contribution due to & Mitigation strategy & Noise & Bias\\
&&$(\Delta g/g)/\sqrt{\mathrm{Hz}}$& $\Delta g/g$\\
\hline
Intrinsic sensitivity &Next generation source~\cite{Rudolph15NJP}&$5.3\cdot 10^{-9}$&$0$\\
Mean-field shift &Tailored expansion and DKC~\cite{Muentinga13PRL,Kovachy15PRL} &$1.5\cdot 10^{-10}$&$6.4\cdot 10^{-11}$\\
Launch velocity &Scatter $70\, \upmu\mathrm{m}/\mathrm{s}$, stability $15\, \upmu\mathrm{m}/\mathrm{s}$~\cite{Louchet11NJP}&$1.5\cdot 10^{-12}$&$3.1\cdot 10^{-13}$\\
Wavefront quality &$\lambda / 10$ chip-coating, $\oslash = 2$\, cm beam~\cite{Tackmann12NJP} &$6.7\cdot 10^{-10}$&$2.8\cdot 10^{-10}$\\
Self gravity &Detailed modeling of chip mount~\cite{Dagostino11Met}&$1.2\cdot 10^{-12}$&$5\cdot 10^{-10}$\\
Light-shifts &Suppressed in Bragg diffraction~\cite{Giese16PRA}&$1.4\cdot 10^{-12}$&$1.4\cdot 10^{-10}$\\
Magnetic fields &Three-layer magnetic shield~\cite{Milke14RSI}&$1\cdot 10^{-10}$&$2.6\cdot 10^{-10}$\\
\hline
Target estimation &Uncertainty after less than 100\,s&\multicolumn{2}{r|}{$\approx 7.8\cdot 10^{-10}$}\\
\hline
		
\end{tabular}
\caption{Estimates of the major systematic uncertainties in the atom-chip gravimeter. 
As a result, the determination of local gravity with a relative accuracy $\Delta g/g < 1 \cdot 10^{-9}$ 
in less than $100\,\textup{s}$ seems possible~\cite{Abend16PRL}. This table is a reproduction of Table~6.9 in Ref.~\cite{Abend17}.}
\label{tab:futuresystematics}
\end{table}

To get a realistic estimate of the dominant uncertainties for a future experiment, our calculations~\cite{Rudolph15NJP} use a state-of-the-art flux of $10^5$ atoms per second at a repetition rate of 1\,Hz. For a free-fall distance of 1\,cm the free-fall time increases to~$\tau_{\mathrm{ToF}}=135$\,ms. If the needed detection separation time stays at~$\tau_{\mathrm{sep}}>15$\,ms, a maximum pulse separation time of~$T=35\, \mathrm{ms}$ 
remains, which combined with a fourth-order beam splitter leads to the shot-noise 
limited intrinsic sensitivity \cite{Abend16PRL}
\begin{equation}
(\Delta g/g)/\sqrt{\mathrm{Hz}} = 5.3\cdot 10^{-9}\,.
\end{equation} 
The flux of $10^5$ atoms/s achieved in the QUANTUS-2 experiment is sufficient~\cite{Rudolph15NJP} 
to reach this inferred sensitivity and a cycle time of roughly 1\,s. 

In addition, we need to be able to detect atoms at the output ports at the shot noise limit, 
which corresponds to~$4.5$\,mrad for this atom flux and cycle time assuming a contrast of $C=0.7$. 
Suppressing the vibrational background noise is the crucial remaining noise source to be mitigated. 
A state-of-the-art vibration isolation would significantly improve the sensitivity, although maximum performance may only be reached 
at a vibrational quiet site~\cite{Tang14RSI}.
	
The mean-field shift can be relaxed when we first lower the atomic density by a faster spreading of the wave packet 
during the $45$\,ms after release from the trap but before relaunch, and then stop it by DKC~\cite{Muentinga13PRL,Kovachy15PRL}. For the final size of $300$\,$\upmu$m at the first pulse, 
$10^5$ atoms and 1\% splitting-ratio stability, phase shifts introduced by the mean field~\cite{Debs11PRA} 
can be sufficiently suppressed below $\upmu$Gal, while expansion rates corresponding to nK temperatures 
preserving the beam-splitter fidelity are achievable.

Fluctuations in the launch velocity, which cause a bias due to the Coriolis effect 
or gravity gradients~\cite{Louchet11NJP,NJP14}, can be characterized to the required level and optimized by the tested release procedure. 
The measured scattering of 70\,$\upmu$m/s and the stability of 15\,$\upmu$m/s of the launch velocity is sufficient to suppress this shift.

The surface quality of the atom chip is crucial for preserving
the high efficiencies and contrasts obtained for lower and higher-order Bragg diffraction and for Bloch oscillations. 
It must be significantly improved for a device of the next generation. 

Indeed, a residual roughness of $\lambda/10$ typical for a standard mirror is assumed here. 
For a beam with a diameter of 2\,cm the phase shifts resulting from the wavefront curvature are insignificant 
since BECs are smaller, and expand slower compared to thermal clouds~\cite{Tackmann12NJP,Schkolnik15APB}. 
Furthermore, the possibility of analyzing the fringe patterns in the density profiles at the exit ports~\cite{Muentinga13PRL,Sugarbaker13PRL,NJP14} may allow the characterization of systematic errors 
arising from wavefront distortions.

The proximity of the atoms to the chip leads to a bias phase shift caused by the gravitational field~\cite{Dagostino11Met} of the chip. 
A mass reduction of the chip mount by a factor of two, combined with a finite-element analysis of the mass distribution which calculates the self-gravity effect with an accuracy of 10$\%$, is sufficient 
to reach the target level.  

Compared to Raman diffraction the influence of light shifts is reduced in MZIs based on Bragg diffraction. 
Since the two-photon light shift scales~\cite{Giese16PRA} with the third power of the inverse of the atomic velocity, 
it is negligible in the fountain geometry.

Finally a three-layered shield instead of a single-layered one, 
resulting in a residual gradient below $10\pm 3$\,mG/m, should be sufficient to suppress any residual bias~\cite{Milke14RSI}.

\subsection{Very long baseline atom interferometry}
\label{sec:VLBAI}

Apart from high-contrast interferometry, delta-kick collimated BECs with effective temperatures below 1\,nK enable extended 
free-evolution times, which can significantly increase the scale factor $kT^2$ for acceleration measurements.
Ground-based setups require a large vacuum vessel to venture into the regime of seconds~\cite{Kovachy15Nature,Hardman16PRL,Zhou11GRG}. For example, a device with a height of 10\,m implies a total free-evolution time $2T=2.8$\,s when operated in the fountain mode \cite{Hartwig15NJP}.
This scenario would increase the scale factor by 25--300 compared to the current generation of gravimeters.

Indeed, with $10^5$ atoms, first-order Bragg diffraction, and a cycle time of 5.3\,s, the shot noise limit for full contrast 
would be at $0.2\,\mathrm{nm/(s^2}\sqrt{\mathrm{Hz}})$, which is competitive with the superconducting gravimeter 
GWR iOSG~\cite{Prothero68RSI}. In contrast to the latter, VLBAI also provides us with absolute measurements and possible further enhancements via large momentum transfer.

Environmental vibrations impose a typical limit in gravimeters, preventing the utilisation of larger scale factors.
Therefore, a sophisticated vibration isolation, correlation with external sensors, 
or a combination of both is required~\cite{Geiger11NatureComm,Barret15NJP,LeGouet08APB}.

Measurement schemes using gravity gradiometers, or testing the UFF intrinsically suppress 
the impact of vibration noise since the relevant quantity is encoded in the differential acceleration between two atom 
interferometers~\cite{Bonnin13PRA,McGuirk02PRA}. Indeed, for gradiometry, ensembles from {\it two} different sources 
can be injected into interferometers, or two ensembles can be generated from a {\it single} source 
via a large momentum beam splitting process~\cite{Asenbaum17PRL}. The latter approach implies a well-defined distance between the two interferometers which avoids noise contributions from a relative position jitter due to the outcoupling from two different sources and reduces systematic errors. Operating the interferometer with $10^5$ atoms divided into two ensembles, first-order Bragg diffraction in the interferometer, 
a total interferometer time $2T=1$\,s, a cycle time of 4\,s, and a baseline of 5\,m between the interferometers 
would lead to a shot noise limit of $6.3\cdot10^{-10}\,\mathrm{/(s^2}\sqrt{\mathrm{Hz}})$ for gravity gradients. Further improvements are possible by upgrading the first-order Bragg diffraction to large momentum transfer.

A test of UFF with $^{87}$Rb and $^{170}$Yb may lead to the shot noise limit~$0.1\,\mathrm{nm/(s^2}\sqrt{\mathrm{Hz}})$ of measurements of differential accelerations, implying a statistical uncertainty~$4\cdot 10^{-14}$ in the E\"otv\"os parameter after 24\,h of integration, which is competitive with experiments on the ground achieving~$10^{-13}$~\cite{Williams12CQG,Wagner2012CQGtorsion,Mueller12CQG}, and in space 
reaching~$10^{-14}$~\cite{Touboul2017PRLMICROSCOPE}, with the added benefit of different species.
The assumptions are $2\cdot10^5$ ($1\cdot10^5$) atoms, eighth- (fourth-) order Bragg transitions 
at 780\,nm (at 399\,nm) for $^{87}$Rb ($^{170}$Yb), a total free evolution time $2T=2.6$\,s, and a cycle time 
of 12.6\,s~\cite{Hartwig15NJP}.
Here, the transfer functions of the two interferometers will not be ideally matched due to the different wave vectors, 
requiring an auxiliary sensor to suppress vibration noise via cross correlation~\cite{Barret15NJP,barrett2016dual}.

Relevant systematics~\cite{Hartwig15NJP} originates from (i) wavefront errors~\cite{Louchet11NJP,Schkolnik15APB} suppressed by the low and matched expansion rates to be traded off against residual mean-field contributions \cite{Debs11PRA}, 
(ii) magnetic field inhomogeneities affecting $^{87}$Rb~\cite{Gauguet09PRA,PhysRevD.78.122002} and reduced by a magnetic shield, 
(iii) rotations countered by a tip-tilt stage~\cite{Freier16CS,Dickerson13PRL,PhysRevLett.108.090402} and (iv) gravity gradients coupling to the relative displacements of the two elements, 
which have to be assessed with the device itself~\cite{hogan2008light}. In fact, the requirements on the relative position and velocity 
of the two initial wave packets due to gravity gradients can be substantially relaxed thanks to an effective compensation technique 
proposed in Ref. \cite{PRL118}, which has been experimentally demonstrated both 
for UFF tests \cite{PRL120} and gradiometry measurements \cite{PRL119}.

\subsection{Space-borne atom interferometers}
\label{sec:space}

The microgravity environment of a space-borne atom interferometer provides us with access to even longer free-evolution times. Moreover, no seismic noise disturbs the measurement.
Indeed, since both the device and the atomic ensemble are in free-fall, the movement of the atoms with respect to the potentials 
for trapping and DKC is decreased, opening up a different parameter range 
for reducing residual expansion rates and mean-field contributions.
The measurement may even benefit from a much smaller gravitational sag and the absence of a lattice launch. We also have the possibility of a signal modulation to suppress systematics 
if the apparatus is inertial pointing~\cite{Aguilera14CQG,Williams16NJP}.

A test of the UFF in space could profit from all these advantages and go beyond an accuracy of about~$10^{-14}$.
The updated Space-Time Explorer and Quantum Equivalence Principle Space Test~(STE-QUEST) scenario~\cite{Wolf15RM}, 
based on a previous version of a dual-species interferometer 
with $^{87}$Rb and $^{85}$Rb~\cite{Aguilera14CQG,Milke14RSI,Altschul15ASR,Schuldt15EA}, 
proposes a dual-species interferometer with $^{87}$Rb and $^{41}$K and a target uncertainty 
in the E\"otv\"os parameter of $2\cdot10^{-15}$.

This goal assumes $10^6$ atoms of each species, beam splitters based on double Bragg diffraction~\cite{Leveque09PRL,Ahlers16PRL}, 
a total interferometer time of $2T=10$\,s, a cycle time of 20\,s, and a highly elliptical orbit 
for the clock comparison part of the mission with a perigee of $\sim$2500\,km, an apogee of $\sim$33600\,km and an orbital period 
of 10.6\,h. Around perigee, the instrument observes a strong signal, whereas around apogee it almost vanishes. Additional measurements 
in between are utilized for calibration.
The target uncertainty is reached after 1.2\,years, limited by the small part of the orbit close to earth.
A circular orbit in a dedicated mission could reduce this time to a few months.

\acknowledgments
We are most grateful to our colleagues H.~Albers, H.~Ahlers, S.~Arnold, D.~Becker, K.~Bongs, H.~Dittus, H.~Duncker, W.~Ertmer, 
A.~Friedrich, N.~Gaaloul, M.~Gebbe, C.~Gherasim, E.~Giese,  C.~Grzeschik, T.W.~H\"ansch, J.~Hartwig, O.~Hellmig,  
W.~Herr, S.~Herrmann, E.~Kajari, S.~Kleinert, M.~Krutzik, R.~Kuhl, C.~L\"ammerzahl, W.~Lewoczko-Adamczyk, J.~Malcolm, N.~Meyer, 
H.~M\"untinga, R.~Nolte, A.~Peters, M.~Popp, J.~Reichel, L.L.~Richardson, J.~Rudolph, M.~Schiemangk, M.~Schneider, S.T.~Seidel, 
K.~Sengstock, G.~Tackmann, V.~Tamma, T.~Valenzuela, A.~Vogel, R.~Walser, T.~Wendrich, A.~Wenzlawski, P.~Windpassinger, W.~Zeller, 
T.~van Zoest, who have worked with us very closely over the years on the topics presented in these lecture notes.

We acknowledge the support by the CRC 1227 DQmat, the CRC 1128 geo-Q, and the QUEST-LFS. The German Space Agency (DLR) 
with funds provided by the Federal Ministry of Economic Affairs and Energy (BMWi) 
due to an enactment of the German Bundestag under Grant No. DLR 50WM1137 (QUANTUS-IV-Fallturm), and 50WM1641 (PRIMUS-III) was central to our work. We also acknowledge support by the Federal Ministry of Education and Research (BMBF), by ''Wege in die Forschung (II)'' of Leibniz Universit\"at Hannover, and ''Nieders\"achsisches Vorab'' through the ''Quantum- and Nano-Metrology (QUANOMET)'' initiative.

M.A.E. thanks the Center for Integrated Quantum Science and Technology (IQ$^{ST}$) for financial support.
W.P.S.  is  most  grateful  to  Texas A$\&$M  University  for  a  Faculty  Fellowship  at  the  Hagler  Institute  for  
Advanced Study at the Texas A$\&$M University as well as to Texas A$\&$M AgriLife Research. 
F.A.N. acknowledges a generous grant from the Office of 
the Secretary of Defense (OSD) in the Quantum Science and Engineering Program (qSEP).

\bibliography{varenna_lecture}

\end{document}